%% file: main.tex
\RequirePackage{etex}
\documentclass{aa}
\let\linenumbers\nolinenumbers
\usepackage{graphicx}
\usepackage{amsmath}
\usepackage{array}
\usepackage[varg]{txfonts}
\usepackage{enumitem}
\usepackage{hyperref}
\usepackage{xcolor}
\begin{document}

\title{Tracking the evolution of sub-Neptunes}
\subtitle{III. The irradiation dependence of the radius valley}

\author{N.~Greco\inst{\ref{inst:0001},\ref{inst:0002}}\corrauth{nino.greco@inaf.it}
\and A.C.~Lanzafame\inst{\ref{inst:0001},\ref{inst:0002}}\email{a.lanzafame@unict.it} \and I.~Pagano\inst{\ref{inst:0003}}\email{isabella.pagano@inaf.it} \and G.~Scandariato\inst{\ref{inst:0002}}\email{gaetano.scandariato@inaf.it}
      }

   \institute{
Dipartimento di Fisica e Astronomia ``Ettore Majorana'', Universit\`{a} di Catania, Via S. Sofia 64, 95123 Catania, Italy\relax\label{inst:0001} \and INAF - Osservatorio Astrofisico di Catania, Via S. Sofia 78, 95123 Catania, Italy\relax \label{inst:0002} \and INAF, Istituto Nazionale di Astrofisica, Viale del Parco Mellini 84, 00136 Roma, Italy\relax \label{inst:0003}
   }

\date{Received ; accepted }

\input{sections/00_abstract}

\maketitle

\input{sections/01_introduction}

\input{sections/02_sample_model}

\input{sections/03_linear_forms}

\input{sections/04_threshold_forms}

\input{sections/05_age_dependence}

\input{sections/06_planet_mass}

\input{sections/07a_discussion_mechanisms}
\input{sections/07b_discussion_threshold}

\input{sections/07c_discussion_metallicity_limits}

\input{sections/08_conclusions}

\begin{acknowledgements}
This work was partially supported by the European Union -- NextGenerationEU, under the Italian Ministry of University and Research (MUR) National Recovery and Resilience Plan (PNRR), Mission 4, Component 2, Investment 1.1, PRIN 2022, project No. 2022HY2NSX, ``CHRONOS: adjusting the clock(s) to unveil the CHRONO-chemo-dynamical Structure of the Galaxy'' and the European Union's Horizon 2020 research and innovation programme under Grant Agreement No. 824135 (SOLARNET -- Integrating High Resolution Solar Physics).
\end{acknowledgements}

\bibliographystyle{aa}
\bibliography{references}

\appendix

\input{sections/A_fluence_variables}

\input{sections/B_model_details}

\input{sections/C_simulated_catalogues}
\input{sections/D_planet_masses}

\input{sections/E_closed_form}

\end{document}

%% file: sections/00_abstract.tex
\abstract
{The radius valley separating super-Earths from sub-Neptunes is attributed to atmospheric mass loss, but it remains debated whether it is set by the X-ray and extreme-ultraviolet (XUV) irradiation of the planets or by their bolometric irradiation. The first drives photoevaporation, and the second regulates core-powered mass loss.}
{We test whether the fraction of super-Earths, corrected for the selection of the survey, is set by a threshold in the XUV fluence, the XUV energy per unit area that each planet has received over its lifetime.}
{We use $581$ \textit{Kepler} planets around FGK dwarfs with gyrochronological ages and a likelihood that accounts for their detection and confirmation. We compare the XUV fluence, computed from the rotational and activity evolution for the mass and age of each host, with the orbital period, the insolation, and the bolometric fluence.}
{The super-Earth fraction rises across a narrow threshold at an XUV fluence of about $10^{21}$\,erg\,cm$^{-2}$. The XUV fluence describes this fraction better than the insolation and the bolometric fluence. It removes the need for the dependence on host mass found at fixed insolation and requires no further dependence on age. There is tentative evidence that the threshold lies at higher XUV fluence around metal-rich hosts. For FGK hosts, the orbital period is a close proxy for the XUV fluence.}
{The threshold is comparable to the XUV fluence needed to remove the envelope of a rocky core of a few Earth masses. The dependences on host mass and, less significantly, on age are those expected from the XUV irradiation, whereas the bolometric irradiation reproduces the first only if the core masses grow with host mass, a growth that the super-Earth radii do not show. The data thus tentatively favour photoevaporation over core-powered mass loss.}

\keywords{planets and satellites: physical evolution -- planets and satellites: atmospheres -- methods: statistical}

%% file: sections/01_introduction.tex
\section{Introduction}\label{sec:intro}

The radii of close-in small planets are bimodally distributed: a deficit of planets between $\sim$$1.5$ and $2\,R_\oplus$, the radius valley, separates super-Earths (SE) from sub-Neptunes (SN) \citep{Fulton2017,FultonPetigura2018,VanEylen2018}. The valley is commonly attributed to thermally driven atmospheric mass loss, through two mechanisms that respond to different components of the stellar irradiation. In photoevaporation, the outflow is driven by the X-ray and extreme-ultraviolet (XUV) radiation of the host \citep{OwenWu2013,LopezFortney2013,OwenWu2017}. In a solar-mass star this emission remains saturated for the first $10$--$300$\,Myr, depending on the initial rotation \citep{Tu2015}, and then declines at a rate that depends on stellar mass \citep{Johnstone2021}, more slowly in the extreme ultraviolet than in X-rays \citep{KingWheatley2021}. In the energy-limited approximation, a planet can lose its envelope when the XUV energy it absorbs, integrated over time and weighted by the escape efficiency, becomes comparable to the gravitational binding energy of the envelope \citep{OwenWu2017}. In core-powered mass loss, the outflow is powered by the cooling of the planetary core and regulated by the equilibrium temperature, and hence by the insolation \citep{Ginzburg2018,GuptaSchlichting2019,GuptaSchlichting2020}.

Both mechanisms also depend on planetary core mass, while the irradiation history depends on the properties of the host star. Stellar mass and metallicity may therefore affect the observed planet population through both stellar irradiation and planetary properties, complicating the distinction between atmospheric-loss mechanisms \citep{GuptaSchlichting2020,Petigura2022}. At orbital periods below $10$\,d, both SEs and SNs are more common around metal-rich hosts, whereas at $10$--$100$\,d only the SNs are \citep{Petigura2018}. \citet{OwenMurrayClay2018} attributed this excess of SNs to more massive cores and to the slower photoevaporation of metal-enriched envelopes.

The dependence of the valley on orbital period, insolation, and stellar mass has been used to distinguish between the two mechanisms. The valley radius decreases with orbital period and increases with stellar mass \citep{FultonPetigura2018,Petigura2022,HoVanEylen2023}. Core-powered mass loss has been found consistent with the shift and shallowing of the valley over Gyr timescales \citep{Kamulali2026} and favoured by a valley that is present among magnetically quiet hosts and not systematically deeper among the most active ones \citep{Chen2026}. By contrast, the shallower valley around low-mass hosts may reflect a scatter in the XUV evolution of the hosts, a range of core compositions, or collisions \citep{Ho2024}. 
A complementary approach is to consider the irradiation accumulated over the lifetime of each system. \citet{McDonald2019} applied this approach to the sub-Neptune desert, the scarcity of SNs at the highest insolations, and found that the insolation distributions of SNs peak at different values around hosts of different spectral types, but peak at nearly the same value when expressed in terms of accumulated X-ray fluence. This convergence arises because the higher and longer activity saturation of lower-mass stars partly compensates for their lower luminosities. Their result suggests that the cumulative high-energy irradiation may provide a more physically meaningful description of atmospheric loss than insolation alone. Whether the same holds for the radius valley, and particularly for the intrinsic relative abundance of SEs and SNs, remains to be established.

These studies rely on the detected planets, in some cases weighted by the inverse of their transit and detection probabilities \citep{FultonPetigura2018,Petigura2022}. The features of the radius distribution they measure, and the trends of these features with stellar and planetary properties, therefore depend on the treatment of the survey selection. This matters most at low irradiation, where the SEs lie at long orbital periods and the completeness of the Kepler catalogue is lowest \citep{Thompson2018}.

The population also evolves with age: the ratio of SEs to SNs increases with host age \citep{Berger2020,Sandoval2021,Chen2022,PaperII}, the valley lies at smaller radii among younger planets \citep{David2021} and is not fully established below $\sim$$3$\,Gyr \citep{Chen2026}, and short-period planets above the valley are more frequent around stars younger than $1$\,Gyr than around older field stars \citep{Christiansen2023,Vach2024,Fernandes2025}.

In the first two papers of this series, we derived gyrochronological ages for planet hosts \citep[][hereafter Paper~I]{PaperI} and measured the radius valley and the SE fraction as a function of age \citep[][hereafter Paper~II]{PaperII}. Paper~II modelled the radius distribution as a mixture of a SE and a SN component, with a likelihood conditioned on the detection and confirmation of each planet, and found that the SE weight depends on orbital period and increases with age up to $\sim$$2$\,Gyr. Here we use the same sample and likelihood to test whether the SE weight is better described by the XUV fluence accumulated over the activity evolution of each host than by the bolometric irradiation, and whether the accumulated fluence accounts for its dependence on age.

This paper is organised as follows. Section~\ref{sec:data} presents the sample, the irradiation variables, and the model. Sections~\ref{sec:linear} and~\ref{sec:law} compare the irradiation variables, Sect.~\ref{sec:age} examines the dependence on age, and Sect.~\ref{sec:mass} relates the results to the planet masses. Sections~\ref{sec:disc} and~\ref{sec:conclusions} discuss and summarise the results.

%% file: sections/02_sample_model.tex
\section{Sample, irradiation variables, and model}\label{sec:data}

\subsection{Sample}\label{sec:sample}
We use the reference sample $\mathcal{D}_3$ of Paper~II, comprising $581$ \textit{Kepler} Objects of Interest (KOIs) from DR25 \citep{Thompson2018} around $404$ hosts with the DR25 per-star detection metrics \citep{BurkeCatanzarite2017,BurkeCatanzarite2017b}. These metrics allow the detection efficiency to be evaluated for each planet. The sample consists of planets with radii of $0.5$--$4\,R_\oplus$, orbital periods up to $100$\,d, and reliability tiers $\le2$ for both radius and period on the four-tier scale of Paper~II (with tier 1 indicating the highest reliability). Their FGK hosts have de-reddened \textit{Gaia} DR3 \citep{GaiaDR3} colours $(G_{\rm BP}-G_{\rm RP})_0$ of $0.65$--$1.90$ and gyrochronological ages from Paper~I with reliability tier $\le2$, between $0.32$ and $3.98$\,Gyr. 

Of the $581$ planets, $377$ orbit K and late-G hosts, defined by $(G_{\rm BP}-G_{\rm RP})_0>0.95$, and $204$ orbit F and G hosts. A total of $318$ planets have at least one other KOI around the same host that is confirmed, hereafter referred to as a \emph{confirmed sibling}. The host metallicities are those adopted in Paper~I, with a median of $+0.04$ and a $5$th--$95$th percentile range from $-0.22$ to $+0.24$ across $\mathcal{D}_3$. The stellar parameters are described in Paper~I, and the planetary parameters, selection criteria, and database in Paper~II.

\subsection{Irradiation and fluence}\label{sec:axes}
The insolation of a planet is $F=(L/L_\odot)\,(a_p/{\rm au})^{-2}$, in units of the solar constant $F_\oplus=1.361\times10^{6}$\,erg\,s$^{-1}$\,cm$^{-2}$, where $L$ is the bolometric luminosity of the host, from Eq.~\eqref{eq:lbol_closed}, and $a_p$ the semi-major axis. The bolometric fluence received up to age $t$ is
\begin{equation}\label{eq:fbol}
\mathcal{F}_{\rm bol}(t)=F\,F_\oplus\int_0^{t}\frac{L_{\rm M}(t')}{L_{\rm M}(t)}\,{\rm d}t',
\end{equation}
where $L_{\rm M}(t)$ is the main-sequence luminosity at age $t$ of a star with the mass of the host, obtained from the MESA Isochrones and Stellar Tracks \citep[MIST;][]{Dotter2016,Choi2016} at solar metallicity (Appendix~\ref{app:axes}). The ratio $L_{\rm M}(t')/L_{\rm M}(t)$ describes the luminosity evolution relative to the present age, while the observed luminosity sets its normalisation. We use $\mathcal{F}_{\rm bol}$ as the bolometric counterpart of the accumulated XUV fluence, 
to test atmospheric mass loss regulated by bolometric irradiation over gigayear timescales 
\citep{GuptaSchlichting2020}. The accumulated X-ray fluence is
\begin{equation}\label{eq:fx}
\mathcal{F}_{\rm X}(t)=F\,F_\oplus\int_0^{t}R_{\rm X}(t')\,\frac{L_{\rm M}(t')}{L_{\rm M}(t)}\,{\rm d}t',
\end{equation}
where $R_{\rm X}=L_{\rm X}/L$ is the ratio of the X-ray luminosity $L_{\rm X}$ in the $0.1$--$2.4$\,keV band to the bolometric luminosity, computed from the rotation period that the model of Paper~I gives for a star of the same mass at each age, with the rotation--activity relation of \citet{Johnstone2021} (Appendix~\ref{app:axes}).

The accumulated XUV fluence $\mathcal{F}_{\rm XUV}$ adds to the X-ray emission the extreme-ultraviolet emission at $10$--$92$\,nm. It is computed by replacing $R_{\rm X}$ in Eq.~\eqref{eq:fx} with $R_{\rm XUV}=(L_{\rm X}+L_{\rm EUV})/L$, where the extreme-ultraviolet luminosity $L_{\rm EUV}$ is obtained from $L_{\rm X}$ with the relations of \citet{Johnstone2021} (Appendix~\ref{app:axes}). Since photoevaporation is driven by the whole XUV emission \citep{OwenWu2017}, we use $\mathcal{F}_{\rm XUV}$ instead of $\mathcal{F}_{\rm X}$ in the following sections.

To assess the sensitivity of the results to the adopted activity evolution, we also compute the XUV fluence using the open-cluster calibration of \citet{Jackson2012}. In this prescription, $R_{\rm X}$ remains at a constant saturated level up to a saturation turn-off age and subsequently declines as a power law. We denote the resulting fluence by $\mathcal{F}^{\rm J}_{\rm XUV}$. The two XUV prescriptions differ only in the evolution of $R_{\rm X}(t)$; they use the same bolometric luminosity evolution, EUV scaling relations, and integration procedure (Appendix~\ref{app:axes}). Since the accumulated fluences depend on stellar age, they are evaluated for each age draw used to marginalise the likelihood over the age uncertainty.

\subsection{Likelihood and super-Earth weight}\label{sec:model}
We adopt the per-planet likelihood developed in Paper~II, in which the distribution of $\log_{10}R_p$ is represented by a mixture of two Gaussian components corresponding to SEs and SNs, with parameters common to all planets. The likelihood accounts for the \emph{selection} of each planet, the probability of detecting and confirming it as a function of its true radius, and is normalised over the radius range of the sample. It is marginalised over the uncertainty in host age using $200$ age draws.

All models considered here share this likelihood and differ only in the SE mixture weight $w_i$, defined as the probability that planet $i$ belongs to the SE component, given its properties other than radius. The average of these weights over a specified population gives the intrinsic SE fraction predicted by the model, before applying survey selection effects.

In Paper~II, the weight was modelled as
\begin{equation}\label{eq:wpaperii}
{\rm logit}\,w_i =
{\rm logit}\,f_{\rm SE}^{\rm m}(t)
+\beta_P(\log_{10}P_i-\log_{10}P_{\rm ref})
+\gamma g_i,
\end{equation}
where ${\rm logit}\,w=\ln[w/(1-w)]$ and $f_{\rm SE}^{\rm m}(t)$ is the SE weight at age $t$ for a planet without a confirmed sibling at the reference period $P_{\rm ref}=8.7$\,d, the median of $\mathcal{D}_3$. The coefficient $\beta_P$ describes the dependence on orbital period, while $\gamma$ accounts for the presence of confirmed siblings, with $g_i=1$ for planets with at least one confirmed sibling and $g_i=0$ otherwise.

\subsection{Fits and comparisons}\label{sec:fits}

We fit the models of the SE weight of Sects.~\ref{sec:linear}--\ref{sec:age} to the $581$ planets of $\mathcal{D}_3$. We compare them using the difference of their log-likelihoods at the maximum a posteriori (MAP), $\Delta\ln\mathcal{L}$, and the Bayesian information criterion, ${\rm BIC}=n_{\rm par}\ln N-2\ln\mathcal{L}$, where $n_{\rm par}$ is the number of free parameters and $N=581$ the number of planets. 
Using the $404$ independent hosts instead of the planets in the BIC penalty does not change the ranking of the threshold forms of Sect.~\ref{sec:law} (Appendix~\ref{app:fits}). We sample the parameter posteriors with \texttt{emcee} \citep{ForemanMackey2013}, as detailed in Appendix~\ref{app:fits}. The models are further assessed using planets held out of the fit (Appendix~\ref{app:tests}) and simulated catalogues subject to the same selection effects (Appendix~\ref{app:mocks}).

Unless otherwise stated, parameter uncertainties are given by the 16th--84th percentile intervals of the posterior. Values quoted without uncertainties refer to the MAP, while intervals for individual coefficients evaluated at the MAP are obtained from the profile likelihood at $\Delta\ln\mathcal{L}=0.5$.

%% file: sections/03_linear_forms.tex
\section{Linear forms of the super-Earth weight}\label{sec:linear}

We first compare models in which the logit of the SE weight depends linearly on orbital period, insolation, or accumulated XUV fluence, expressed on logarithmic scales. We also test whether the host colour or colour subset (K and late-G versus F and G; Sect.~\ref{sec:sample}) provides additional information. We modify only the period term of Eq.~\eqref{eq:wpaperii}, keeping $f_{\rm SE}^{\rm m}(t)$ constant within the three age intervals of Paper~II, $0.1$--$1$, $1$--$2$, and $2$--$4$\,Gyr, and compare the resulting \emph{linear forms} at the MAP. The period dependence is strongly supported: removing it decreases the log-likelihood by $23.8$ ($\Delta{\rm BIC}=41.2$). Replacing $\log_{10}P$ with $\log_{10}F$, without changing the number of parameters, decreases the log-likelihood by $5.7$ ($\Delta{\rm BIC}=11.4$ in favour of the period). Adding a linear term in insolation, host colour, or colour subset to the period model improves the log-likelihood by at most $0.11$.

A dependence on host colour appears when insolation is used instead of period. At fixed insolation, the SE weight of K and late-G hosts exceeds that of F and G hosts by $0.15\pm0.05$, averaged over the planets of $\mathcal{D}_3$. This difference can arise from the correlation between orbital period, insolation, and host properties. At a given period, planets around K and late-G hosts receive a median insolation lower by a factor of $2.3$ ($0.37$\,dex) than those around F and G hosts. Conversely, at a given insolation, planets around the less luminous hosts generally have shorter orbital periods. Since the SE weight decreases with increasing period, a dependence on period alone can therefore produce an apparent excess of SEs around the redder hosts at fixed insolation.

Replacing the period with the accumulated XUV fluence lowers the log-likelihood by $0.6$. When a linear metallicity term is included in both models, the XUV-fluence model has a log-likelihood higher by $0.3$. Thus, within these linear forms, the data do not distinguish between orbital period and accumulated XUV fluence as predictors of the SE weight.

%% file: sections/04_threshold_forms.tex
\section{Threshold forms of the super-Earth weight}\label{sec:law}

\subsection{The threshold forms}\label{sec:thrforms}
We replace $f_{\rm SE}^{\rm m}(t)$ and the period term of Eq.~\eqref{eq:wpaperii} with a threshold in a single irradiation variable, allowing for a metallicity-dependent shift. These \emph{threshold forms} read
\begin{equation}\label{eq:law}
{\rm logit}\,w_i={\rm logit}\,\Theta(h_i+\xi_Z\,\Delta_i)+\gamma\,g_i,
\end{equation}
\begin{equation}\label{eq:W}
\Theta(y)=w_{\rm b}+(1-w_{\rm b})\,\Phi\!\left(\frac{y-y_{\rm c}}{\sigma_y}\right),
\end{equation}
where $\Phi$ is the standard normal cumulative distribution function, so that the weight rises from $w_{\rm b}$ to unity across a transition centred at $y_{\rm c}$. Each threshold form uses one of five irradiation variables, $\log_{10}\mathcal{F}_{\rm XUV}$, $\log_{10}\mathcal{F}^{\rm J}_{\rm XUV}$, $\log_{10}\mathcal{F}_{\rm bol}$, $\log_{10}F$, or $-\log_{10}P$, whose value for planet $i$ is $h_i$. In the metallicity term $\xi_Z\,\Delta_i$, $\xi_Z$ is the metallicity coefficient and $\Delta_i={\rm [Fe/H]}_i-{\rm [Fe/H]}_0$ is the metallicity of the host relative to the median of $\mathcal{D}_3$, ${\rm [Fe/H]}_0=+0.04$. The forms without the metallicity term have $\xi_Z=0$. For the threshold in $\log_{10}\mathcal{F}_{\rm XUV}$, we write $\sigma_{\mathcal{F}}$ for $\sigma_y$, and the argument of $\Theta$ is $\log_{10}\mathcal{F}_Z$, with the XUV fluence scaled by the metallicity term
\begin{equation}\label{eq:fz}
\mathcal{F}_Z=\mathcal{F}_{\rm XUV}\,10^{\xi_Z\Delta_i}.
\end{equation}
The threshold in $\log_{10}\mathcal{F}_{\rm XUV}$ with metallicity is the reference form of this work (Sect.~\ref{sec:lawfit}); we refer to it as \emph{the reference threshold}.

The threshold form admits a population-level interpretation in which planets have different critical values of the irradiation variable $h$, above which they belong to the SE component. These critical values follow a normal distribution in $h$, with median $y_{\rm c}-\xi_Z\Delta_i$ and width $\sigma_y$. For logarithmic irradiation variables, this corresponds to a log-normal distribution of physical thresholds. A negative $\xi_Z$ therefore shifts the threshold towards higher irradiation around metal-rich hosts. In this interpretation, a fraction $w_{\rm b}$ of the population, the \emph{background fraction}, belongs to the SE component independently of $h$. It is a phenomenological parameter rather than a uniquely identified physical population. In particular, because the SE component extends below $1\,R_\oplus$, where survey completeness is low, its inferred value depends on the assumed shape of the component (Appendix~\ref{app:forms}), as already found in Paper~II.

\subsection{Comparison of the irradiation variables}\label{sec:lawfit}

The threshold forms provide better descriptions of the SE weight than the corresponding linear forms. For orbital period, insolation, and XUV fluence, the improvements in log-likelihood are $3.1$, $2.9$, and $4.6$, respectively. Table~\ref{tab:law} compares the threshold forms in the five irradiation variables, with and without a metallicity term. The threshold in $\mathcal{F}_{\rm XUV}$ with metallicity has the lowest BIC and is adopted as the reference threshold; its fitted parameters are given in Table~\ref{tab:params}.

Compared with the three-interval model of Paper~II, which has the same number of parameters, the reference threshold increases the log-likelihood by $8.2$. Of this improvement, $4.3$ comes from the metallicity term (Sect.~\ref{sec:metal}). The remaining improvement is associated with the non-zero background fraction $w_{\rm b}$: with $w_{\rm b}=0$, a logistic transition has a log-likelihood $9.5$ below the reference threshold and $1.3$ below the model of Paper~II (Appendix~\ref{app:forms}). Thus, the data favour a SE weight that approaches a non-zero value at low XUV fluence rather than continuing to decrease. The possible physical origin of this background fraction is discussed in Sect.~\ref{sec:discmass}. The results are relatively insensitive to the choice of $R_{\rm X}$ before $0.1$\,Gyr and the shape of the transition: alternative prescriptions change the log-likelihood by at most $0.4$ at the same number of parameters (Appendices~\ref{app:axes} and~\ref{app:forms}). Appendix~\ref{app:closedform} gives the reference threshold in closed form.

The XUV-fluence threshold provides a better fit than the thresholds in orbital period, insolation, and bolometric fluence, both with and without the metallicity term (Table~\ref{tab:law}). However, orbital period and XUV fluence are strongly correlated in the sample. At the age of each host, $\log_{10}\mathcal{F}_{\rm XUV}$ and $-\frac{4}{3}\log_{10}P$ have a Pearson correlation coefficient of $0.95$. The scatter of $\log_{10}\mathcal{F}_{\rm XUV}$ around a linear relation with $-\frac{4}{3}\log_{10}P$ is only $0.19$\,dex, about $60\%$ of the width $\sigma_{\mathcal{F}}$ of the reference threshold. Consequently, relatively few planets distinguish the two descriptions, and adding $\log_{10}P$ to the reference threshold improves the log-likelihood by only $0.8$ (Table~\ref{tab:variants}). The tests of Appendices~\ref{app:tests} and~\ref{app:mocks}, summarised in Table~\ref{tab:evidence}, support the advantage of XUV fluence over bolometric irradiation, but do not establish a significant advantage over orbital period.

\begin{table}
\caption{Threshold forms of the SE weight in five variables, fitted to all planets of $\mathcal{D}_3$. The forms follow Eqs.~\eqref{eq:law} and~\eqref{eq:W}, without and with the metallicity term, and the last row is the three-interval model of Paper~II (Eq.~\ref{eq:wpaperii}). $\Delta\ln\mathcal{L}$ and $\Delta$BIC are at the MAP and relative to the reference threshold. Positive $\Delta\ln\mathcal{L}$ and negative $\Delta$BIC favour the form in that row. The forms have eight parameters without the metallicity term and nine with it; the model of Paper~II has nine. $\mathcal{F}_{\rm XUV}$ uses the rotational evolution of Paper~I with the rotation--activity relation of \citet{Johnstone2021}, and $\mathcal{F}^{\rm J}_{\rm XUV}$ the cluster calibration of \citet{Jackson2012}. $\xi_Z$ is the metallicity coefficient at the MAP, in units of the threshold variable per dex of [Fe/H], so that its values are not comparable between variables.}
\label{tab:law}
\centering\small\setlength{\tabcolsep}{4pt}
\begin{tabular}{l|cc|ccc}
\hline\hline
 & \multicolumn{2}{|c|}{without metallicity} & \multicolumn{3}{c}{with metallicity}\\
Variable & $\Delta\ln\mathcal{L}$ & $\Delta$BIC & $\Delta\ln\mathcal{L}$ & $\Delta$BIC & $\xi_Z$\\
\hline
$\log_{10}\mathcal{F}_{\rm XUV}$ & $-4.3$ & $+2.1$ & $0$ & $0$ & $-1.09$\\
$\log_{10}\mathcal{F}^{\rm J}_{\rm XUV}$ & $-5.9$ & $+5.4$ & $-2.3$ & $+4.7$ & $-0.91$\\
$-\log_{10}P$ & $-5.1$ & $+3.8$ & $-2.5$ & $+4.9$ & $-0.58$\\
$\log_{10}F$ & $-11.1$ & $+15.8$ & $-4.6$ & $+9.2$ & $-1.73$\\
$\log_{10}\mathcal{F}_{\rm bol}$ & $-12.1$ & $+17.8$ & $-4.1$ & $+8.2$ & $-2.33$\\
\hline
Paper~II, age intervals & $-8.2$ & $+16.5$ & -- & -- & --\\
\hline
\end{tabular}
\end{table}

\begin{table}
\caption{Parameters of the reference threshold of Eqs.~\eqref{eq:law} and~\eqref{eq:W}. Medians of the posterior with 16th--84th percentile intervals, and values at the MAP (Appendix~\ref{app:fits}). The metallicity is relative to ${\rm [Fe/H]}_0=+0.04$, the median of $\mathcal{D}_3$.}
\label{tab:params}
\centering\small\setlength{\tabcolsep}{4pt}\renewcommand{\arraystretch}{1.35}
\begin{tabular}{lcc}
\hline\hline
Parameter & Posterior & MAP\\
\hline
$w_{\rm b}$ & $0.64^{+0.09}_{-0.10}$ & $0.66$\\
$y_{\rm c}=\log_{10}(\mathcal{F}_{\rm c}/{\rm erg\,cm^{-2}})$ & $21.13^{+0.10}_{-0.14}$ & $21.17$\\
$\sigma_{\mathcal{F}}$ [dex] & $0.32^{+0.12}_{-0.10}$ & $0.26$\\
$\xi_Z$ & $-1.17^{+0.40}_{-0.46}$ & $-1.09$\\
$\gamma$ & $-0.72^{+0.31}_{-0.32}$ & $-0.74$\\
\hline
\end{tabular}
\end{table}

\subsection{The metallicity term}\label{sec:metal}
The metallicity coefficient is negative for every variable (Tables~\ref{tab:law} and~\ref{tab:variants}). In the posterior of the reference threshold, $\xi_Z=-1.17^{+0.40}_{-0.46}$: the median threshold moves to higher XUV fluence around metal-rich hosts. The evidence for the term is tentative. A likelihood-ratio test gives it a $p$-value of $0.004$, which rises to $0.018$ when corrected for the other terms tested on the same data (Appendix~\ref{app:metal}), and the tests of Appendix~\ref{app:tests} do not all confirm its gain (Table~\ref{tab:evidence}).

For the same planets, Paper~II found no significant dependence of the SE weight on host metallicity in a model whose mean radii depend on the host. Here, when the mean radii of the SE and SN components depend on host mass and [Fe/H], the metallicity term in the weight $w_i$ increases the log-likelihood by $2.0$ instead of $4.3$ and no longer lowers the BIC; the threshold in $\mathcal{F}_{\rm XUV}$ without the metallicity term then has the lowest BIC of all the forms of this paper (Appendix~\ref{app:metal}).
The metallicity dependence of the SE weight is therefore not uniquely identified: part of the apparent shift in the irradiation threshold can instead be described by a metallicity dependence of the component radii. Importantly, the preference for XUV fluence over the other irradiation variables persists under either description.
With or without the metallicity term, the threshold in $\mathcal{F}_{\rm XUV}$ still describes the weight better than those in the other variables, so that the comparison of the irradiation variables does not depend on this choice. The reference threshold, like the reference model of Paper~II, has mean radii common to all planets, so that the dependence on irradiation is carried by the weight $w_i$ alone.

The metallicities of $\mathcal{D}_3$ span a narrow range, from $-0.22$ to $+0.24$ at the 5th and 95th percentiles, and [Fe/H] correlates with host mass (Pearson coefficient $0.47$), so that the metallicity term can in part stand for a dependence on host mass. For the reference threshold this part is small: with a host-mass term added (Sect.~\ref{sec:host}), $\xi_Z$ changes from $-1.09$ to $-0.68$ at the MAP, within about one standard deviation of its posterior. It is larger for the variables that carry less of the dependence on host mass, for which adding the metallicity term increases the log-likelihood more (Table~\ref{tab:law}). The constraint on $\xi_Z$ comes mostly from the K and late-G hosts and from the planets with periods shorter than $10$\,d. With separate coefficients for the K and late-G hosts, with $(G_{\rm BP}-G_{\rm RP})_0>0.95$, and for the bluer F and G hosts (Sect.~\ref{sec:sample}), $\xi_Z$ is $-1.33^{+0.39}_{-0.58}$ and $-0.19^{+0.65}_{-0.58}$, respectively. Beyond $10$\,d, the metallicity term does not improve the fit.

Neglecting measurement uncertainties in [Fe/H] tends to attenuate the fitted metallicity dependence. When these uncertainties are propagated, $\xi_Z$ becomes more negative, although the size of the correction depends on the assumed population distribution of [Fe/H]. The inferred threshold width and the comparison among irradiation variables remain essentially unchanged (Appendix~\ref{app:metal}).

\subsection{Host colour and host mass}\label{sec:host}
The reference threshold has no term in host colour or mass, and at fixed $\mathcal{F}_{\rm XUV}$ and metallicity the SE weight does not require a term in either: a shift of the threshold for K and late-G hosts or a host-mass term increases the log-likelihood by only $1.3$ and $1.2$ (Table~\ref{tab:variants}). This holds also without the metallicity term, in which case the colour term does not lower the BIC of the threshold in fluence $\mathcal{F}_{\rm XUV}$ and the host-mass term lowers it by only $0.5$. The threshold in insolation $F$, in contrast, requires a host-mass term, which lowers its BIC substantially (Tables~\ref{tab:law} and~\ref{tab:variants}).

The same result appears when the SE weight, instead of having the functional form of Eqs.~\eqref{eq:law} and~\eqref{eq:W}, is left free in five intervals of insolation, of bolometric fluence, or of $\mathcal{F}_Z$, separately for the K and late-G hosts and for the F and G hosts. At fixed insolation $F$ and at fixed bolometric fluence $\mathcal{F}_{\rm bol}$, its average over the planets of an interval is higher for K and late-G hosts than for F and G hosts by $+0.15^{+0.06}_{-0.08}$ and $+0.15^{+0.06}_{-0.07}$, respectively. At fixed $\mathcal{F}_Z$, the difference falls to $+0.07^{+0.06}_{-0.05}$: 
the residual dependence on host colour is substantially reduced and becomes statistically consistent with zero when the population is described by $\mathcal{F}_{\rm XUV}$.
For the same planets and intervals, the reference threshold gives differences of $+0.04\pm0.01$ and $-0.01\pm0.01$, both consistent with the measured ones, and the counted fractions of the two groups of hosts follow the same curve along $\mathcal{F}_Z$ (Sect.~\ref{sec:ppc} and Fig.~\ref{fig:ppc}).

\begin{figure}\centering
\includegraphics[width=0.85\columnwidth]{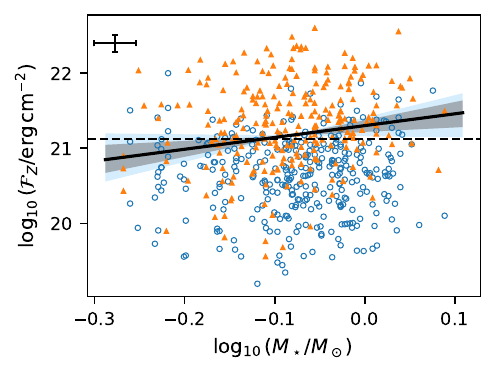}
\caption{Planets of $\mathcal{D}_3$ in the plane of host mass and fluence $\mathcal{F}_Z$ (Eq.~\ref{eq:fz}). Orange triangles: SEs. Open blue circles: SNs. Solid line: threshold in $\mathcal{F}_{\rm XUV}$ with the host-mass term of Eq.~\eqref{eq:hostmass}, at the MAP; its slope is $-\xi_M$. Grey and light blue bands: the same threshold over the $68\%$ and $90\%$ intervals of $\xi_M$, respectively. Dashed line: reference threshold, Eqs.~\eqref{eq:law} and~\eqref{eq:W}, at the $y_{\rm c}$ of Table~\ref{tab:params}. Cross: median uncertainties of the planets.}\label{fig:massplane}
\end{figure}

The dependence on host colour at fixed insolation can be expressed as a dependence of the threshold on host mass, which is closely related to main-sequence colour and is also the variable used to compute $R_{\rm X}$. We quantify this dependence by fitting the threshold form of Eqs.~\eqref{eq:law} and~\eqref{eq:W} with
\begin{equation}\label{eq:hostmass}
h_i=\log_{10}F+\xi_M\log_{10}(M_\star/M_{\star,0}),
\end{equation}
where $M_{\star,0}=0.84\,M_\odot$ is the median host mass of the planets in $\mathcal{D}_3$. Including the metallicity term of Eq.~\eqref{eq:law}, we obtain $\xi_M=-3.26^{+1.05}_{-1.09}$, corresponding to a threshold insolation that increases as $F_{\rm thr}\propto M_\star^{3.26}$ at fixed metallicity. Replacing $F$ with $\mathcal{F}_{\rm XUV}$ in Eq.~\eqref{eq:hostmass} gives $\xi_M=-1.57^{+0.97}_{-0.86}$, with zero within the $90\%$ interval (Fig.~\ref{fig:massplane}), and the host-mass term raises the BIC by $4.0$, whereas it lowers that of the threshold in $F$ by $2.1$ (Table~\ref{tab:variants}). Thus, the XUV fluence removes the need for the host-mass term that the insolation requires. Appendix~\ref{app:host} compares the host-mass dependences implicit in $P$, $\mathcal{F}^{\rm J}_{\rm XUV}$, $\mathcal{F}_{\rm XUV}$, and $\mathcal{F}_{\rm bol}$ with the fitted values of $\xi_M$.

At fixed orbital period $P$, the planets of K and late-G hosts receive less irradiation than those of F and G hosts. In the median, their insolation $F$ is lower by a factor of $2.3$ ($0.37$\,dex), but their accumulated XUV fluence $\mathcal{F}_{\rm XUV}$ by a factor of only $1.6$ ($0.20$\,dex). Redder hosts emit a larger fraction of their luminosity at high energies and remain saturated for longer, which offsets part of their lower bolometric luminosity. With the cluster calibration, which \citet{McDonald2019} used for the lifetime X-ray luminosity of FGK hosts, the two groups of hosts differ by a factor of only $1.08$ in $\mathcal{F}^{\rm J}_{\rm XUV}$.

\subsection{Counted fractions and radius components}\label{sec:ppc}

We compare the reference threshold with the counted SE fraction $f_{\rm SE}^{\rm c}$ defined in Paper~II. A planet is classified as a SE if its radius lies below the period-dependent valley locus of Paper~II
\begin{equation}\label{eq:locus}
\log_{10}(R_{\rm gap}/R_\oplus)=0.329-0.096\log_{10}(P/{\rm d}),
\end{equation}
and as a SN otherwise. The counted fraction is the proportion of observed planets with radii between $1$ and $4\,R_\oplus$ classified as SEs. Unlike the intrinsic mixture weight, it depends on survey selection effects.

For each planet, the fitted weight $w_i$ and the SE and SN radius components define a distribution of true radii. Combining this distribution with the detection and confirmation probabilities and the measurement uncertainty gives the probability $q_i$ that the planet is observed within $1$--$4\,R_\oplus$, and the joint probability $p_i$ that it is observed in this range and classified as a SE. The expected counted fraction in an interval of $\mathcal{F}_Z$ is then $\sum_i p_i/\sum_i q_i$. This prediction follows from the reference threshold, which is fitted to the measured radii rather than to the counted fractions.

\begin{figure}\centering
\includegraphics[width=\columnwidth]{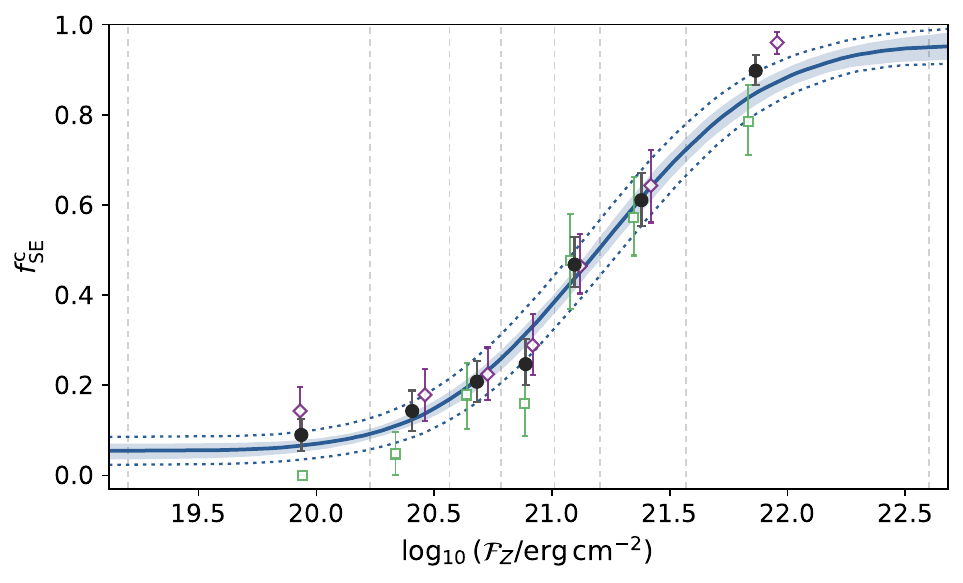}
\caption{Counted SE fraction $f_{\rm SE}^{\rm c}$ in seven intervals of equal population of $\log_{10}\mathcal{F}_Z$ at the age of each host, with $\xi_Z$ at its posterior median (Table~\ref{tab:params}). Black circles: all hosts. Open green squares: F and G hosts. Open purple diamonds: K and late-G hosts. Error bars: 16th--84th percentile intervals from a bootstrap over hosts; the first interval of F and G hosts contains no SE. Solid line and shaded band: fraction expected from the reference threshold, median and 16th--84th percentile range over $300$ samples of the posterior. Dotted lines: the same range combined in quadrature with the binomial scatter of a fraction counted on the $77$ planets of an interval. Dashed vertical lines: interval edges.}\label{fig:ppc}
\end{figure}

Figure~\ref{fig:ppc} compares the predicted and observed counted fractions in seven equal-population intervals of $\log_{10}\mathcal{F}_Z$. They agree within $1.3$ standard deviations in every interval (Appendix~\ref{app:tests}). The two host-colour subsets also follow the same predicted relation (Fig.~\ref{fig:ppc}; Appendix~\ref{app:tests}).

\begin{figure*}\centering
\includegraphics[width=\textwidth]{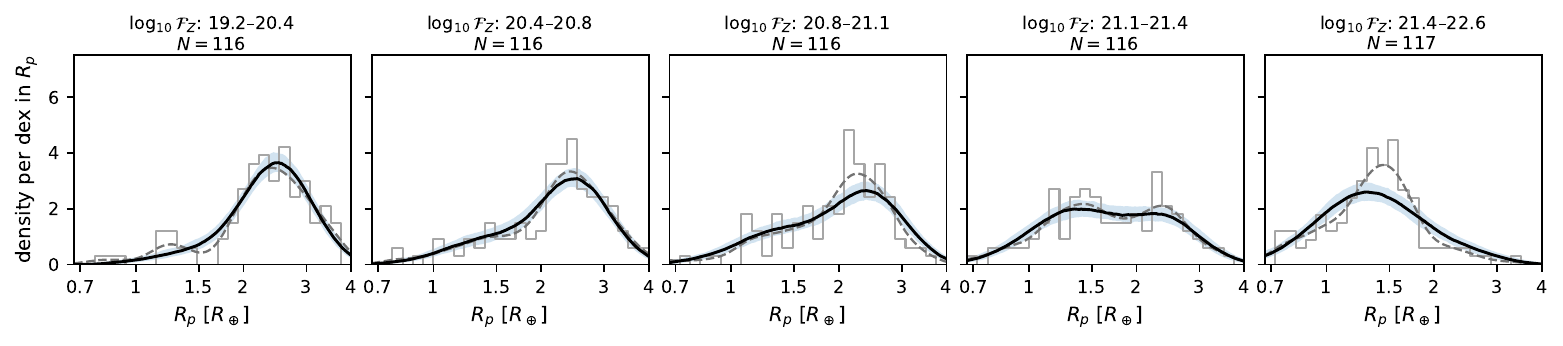}
\caption{Radii of the planets of $\mathcal{D}_3$ in five intervals of equal population of $\log_{10}\mathcal{F}_Z$ at the age of each host, with $\xi_Z$ at its posterior median (Table~\ref{tab:params}); the interval edges and the number of planets are given above each panel. Grey histogram: observed radii, as a density per dex normalised to the planets of the interval; three planets below $0.68\,R_\oplus$ lie outside the axis range. Black line and blue band: median and 16th--84th percentile range of the distribution expected from the reference threshold for the same planets, including their selection. They are obtained from $300$ catalogues simulated from samples of the posterior, smoothed with a Gaussian kernel of $0.045$\,dex. Grey dashed curve: the same kernel applied to the observed radii.}\label{fig:radius1d}
\end{figure*}

The reference threshold assumes that the means and widths of the SE and SN components are the same at every XUV fluence, so that only the weight varies. We fitted the component means separately in five intervals of equal population of $\log_{10}\mathcal{F}_Z$ (Fig.~\ref{fig:radius1d}). 
No significant variation is detected relative to the common values
$1.10^{+0.05}_{-0.06}\,R_\oplus$ for the SE component ($p=0.47$) and $2.46\pm0.05\,R_\oplus$ for the SN component ($p=0.63$). Freeing both means and widths in each interval does not improve the fit significantly ($p=0.21$; Appendix~\ref{app:tests}).

%% file: sections/05_age_dependence.tex
\section{The age dependence at fixed high-energy fluence}\label{sec:age}

Paper~II measured an increase of the SE weight with host age. The reference threshold has no age parameter, but the accumulated XUV fluence $\mathcal{F}_{\rm XUV}$ grows with age, so that the threshold produces an age dependence of its own. Averaged over the age intervals of Paper~II, its $w_i$ agrees with the weight of Paper~II above $1$\,Gyr and is higher below $1$\,Gyr, where only $59$ planets constrain it. In this section, we test whether the data require an age dependence beyond that of $\mathcal{F}_{\rm XUV}$.

We first test whether the thresholds require an age dependence beyond that of their variable. To this end, we add to the argument of $\Theta$ in Eq.~\eqref{eq:law} the term $\xi_t\log_{10}(t/{\rm Gyr})$, where $t$ is the host age and $\xi_t$ a free parameter. For the reference threshold, $\xi_t=-0.11^{+0.26}_{-0.25}$, consistent with zero, providing no evidence for an additional smooth age dependence beyond that already encoded in $\mathcal{F}_{\rm XUV}$.
For the thresholds in $F$ and in $\mathcal{F}_{\rm bol}$, $\xi_t=+0.53\pm0.26$ and $-0.43^{+0.27}_{-0.31}$, which differ from zero by $1.6\sigma$ and $1.5\sigma$, respectively (Appendix~\ref{app:age}). The insolation $F$, which does not change with age, thus leaves an increase of the SE weight with age. 
The negative coefficient for $\mathcal{F}_{\rm bol}$ instead suggests that its accumulated age dependence may be too strong. Neither trend is statistically significant.

To test for a residual age dependence that need not follow a smooth logarithmic trend, we introduce an age-interval-dependent shift $\delta_m$ in the logit of the reference threshold,
\begin{equation}\label{eq:ageshift}
{\rm logit}\,w_i={\rm logit}\,\Theta(h_i+\xi_Z\,\Delta_i)+\gamma\,g_i+\delta_m,
\end{equation}
where $m$ denotes the age interval of each age draw, $<1$, $1$--$2$, or $2$--$4$\,Gyr. We set $\delta_{<1}=0$ for ages below $1$\,Gyr and fit $\delta_{1\text{--}2}$ and $\delta_{2\text{--}4}$ together with the other model parameters. The resulting shifts are $\delta_{1\text{--}2}=1.42^{+0.67}_{-0.64}$ and $\delta_{2\text{--}4}=0.80^{+0.58}_{-0.55}$, improving the log-likelihood by $2.7$. In $19$ of $200$ catalogues simulated without these shifts ($9.5\%$), the improvement is at least as large. Moreover, the BIC favours the reference threshold without additional age terms by $\Delta{\rm BIC}=7.3$. The tests of Appendix~\ref{app:tests} lead to the same conclusion (Table~\ref{tab:evidence}). Thus, the data allow a residual age dependence, but do not require it.

If real, this residual would correspond to an increase in the SE weight near $1$\,Gyr that is steeper than predicted by the accumulated XUV fluence. This behaviour is also suggested by the counted SE fractions: among the $55$ planets with radii of $1$--$4\,R_\oplus$ younger than $1$\,Gyr, $f_{\rm SE}^{\rm c}$ is lower than the value expected from the reference threshold by $0.07^{+0.06}_{-0.05}$, while the counted fractions of older planets lie above the prediction (Appendix~\ref{app:age}). The tentative residual persists when the cluster-calibrated fluence $\mathcal{F}^{\rm J}_{\rm XUV}$ is used and is measured only in the K and late-G hosts, although the two host-colour subsets do not differ significantly (Appendix~\ref{app:age}).

%% file: sections/06_planet_mass.tex
\section{Dependence of the threshold on planet mass}\label{sec:mass}

\subsection{The observed mass--fluence boundary}
\label{sec:massobs}

Equations~\eqref{eq:law} and~\eqref{eq:W} assign the same threshold fluence to all the planets of hosts of a given metallicity (Sect.~\ref{sec:thrforms}). If the envelopes are removed by energy-limited escape, the threshold is also expected to depend on the core mass $M_{\rm c}$, because the binding energy of the envelope increases with core mass.
Since core masses are not directly measured, we use the available total planet masses as proxies, recognising that the approximation depends on the envelope mass fraction.

We consider the $121$ planets of the extended sample $\mathcal{E}$ of Paper~II with measured masses and implied bulk densities between $0.05$ and $15$\,g\,cm$^{-3}$ (Appendix~\ref{app:mass}). This sample includes wider reliability tiers and ages beyond $4$\,Gyr compared with $\mathcal{D}_3$. Of these planets, $29$ lie below the period-dependent valley locus of Paper~II (Eq.~\ref{eq:locus}).

\begin{figure}\centering
\includegraphics[width=0.85\columnwidth]{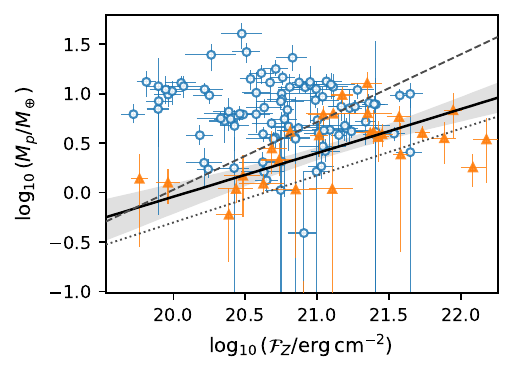}
\caption{Planets of $\mathcal{E}$ with measured masses in the plane of fluence $\mathcal{F}_Z$ (Eq.~\ref{eq:fz}) and planet mass $M_p$. Orange triangles: SEs. Open blue circles: SNs. Error bars: uncertainties of the mass, and of the fluence from the insolation, the age, and the metallicity of the host. Solid line and grey band: median and 16th--84th percentile range of the boundary of Eq.~\eqref{eq:massboundary}. Dashed and dotted lines: cosmic shorelines of \citet{ZahnleCatling2017} and \citet{MeniGallardo2026}, respectively, placed in this plane as described in Appendix~\ref{app:shoreline}.}\label{fig:planetmass}
\end{figure}

In the plane of planet mass $M_p$ and fluence $\mathcal{F}_Z$, SEs have, on average, higher fluences and lower mass than SNs (Fig.~\ref{fig:planetmass}). We describe the probability of belonging to the radius-defined SE class with a logistic function of $\log_{10}\mathcal{F}_Z$ and of the true $\log_{10}M_p$, accounting for the uncertainties of the measured masses (Appendix~\ref{app:mass}). The empirical boundary at which the two classes have equally fitted probabilities is
\begin{equation}\label{eq:massboundary}
\log_{10}\mathcal{F}_Z=h_M+a_M\log_{10}\!\bigl(M_p/2.5\,M_\oplus\bigr),
\end{equation}
where $h_M=21.00^{+0.16}_{-0.14}$ is the value of $\log_{10}(\mathcal{F}_Z/{\rm erg\,cm^{-2}})$ on the boundary at $2.5\,M_\oplus$ and $a_M=2.3^{+0.8}_{-0.5}$ is the slope. 
The constraint is limited by the small number of SEs and by the precision of the masses: $43$ of the $121$ measurements have significance below $3\sigma$.
The fitted slope becomes shallower when only more precise masses are considered (Appendix~\ref{app:mass}).

Because the SE and SN classes are defined by radius, and planet mass correlates with radius, a sloping boundary may arise without any additional dependence of atmospheric loss on mass. We test this by redistributing the measured masses, together with their uncertainties, among planets of similar radius, using $15$ equal-population intervals in $\log_{10}R_p$. Across $1000$ such catalogues, the slope at the MAP is $2.47^{+0.36}_{-0.26}$, and $75\%$ of the catalogues yield a slope at least as large as the observed value of $2.28$. The measured boundary is therefore consistent with that expected from the mass--radius relation and the dependence of the SE fraction on $\mathcal{F}_Z$, without requiring an independent mass dependence. Consequently, the slope $a_M$ does not discriminate between atmospheric-loss mechanisms, and we do not include the mass boundary in the SE-weight model.

For comparison, energy-limited photoevaporation at constant efficiency predicts a threshold fluence scaling approximately as $M_{\rm c}^{a}$, with $a$ between $1.2$ and $2$, depending on the envelope mass fraction and the radii relevant to the binding energy and absorption cross-section (\citealt{OwenWu2017}; \citealt{Mordasini2020}; Appendix~\ref{app:massslopes}). Core-powered mass loss instead depends on bolometric irradiation \citep{GuptaSchlichting2019,GuptaSchlichting2020}. In its Bondi-limited regime, which applies at periods longer than about $8$\,d for the initial envelope fractions adopted by \citet{GuptaSchlichting2019}, the threshold insolation scales as $(M_{\rm c}/R_{\rm c})^4$ \citep{GuptaSchlichting2019,GuptaSchlichting2020}, or approximately $M_{\rm c}^{2.9}$ for rocky cores with $R_{\rm c}\propto M_{\rm c}^{0.27}$ \citep{Valencia2006,Zeng2016}.

Empirical cosmic-shoreline relations provide another comparison. The cosmic shoreline separates the bodies with an atmosphere from those without in the plane of XUV fluence and escape velocity $v_{\rm esc}$. The scalings with escape velocity, $\mathcal{F}_{\rm XUV}\propto v_{\rm esc}^{4}$ and $\mathcal{F}_{\rm XUV}\propto v_{\rm esc}^{5.77}$, reported by \citet{ZahnleCatling2017} and \citet{MeniGallardo2026}, respectively, correspond to slopes $a$ of $1.46$ and $2.11$ for rocky cores. Normalised to the XUV fluence of the Earth, they lie at $\log_{10}(\mathcal{F}_{\rm XUV}/{\rm erg\,cm^{-2}})=20.5$ and $21.5$ at $2.5\,M_\oplus$, respectively, bracketing $h_M$. However, adopting the observed radius of a SN instead of the rocky core radius lowers these values by $0.6$--$0.8$\,dex (Appendix~\ref{app:shoreline}). The comparison is therefore illustrative and depends strongly on which radius is used to define the escape velocity.

\subsection{Core-mass constraints from the threshold width}
\label{sec:masswidth}

Although the observed mass--fluence boundary does not independently constrain the core-mass dependence, the width $\sigma_{\mathcal{F}}$ of the reference threshold provides a conditional constraint on the spread of core masses near the transition. If the physical threshold scales as $\mathcal{F}_{\rm thr}\propto M_{\rm c}^{a}$, a dispersion $\sigma_{M_{\rm c}}$ in $\log_{10}M_{\rm c}$ contributes a dispersion $a\,\sigma_{M_{\rm c}}$ to the logarithmic threshold. Other differences among planets, including their envelope mass fractions, may contribute additional scatter.

The measured width $\sigma_{\mathcal{F}}$ combines the intrinsic spread $\sigma_{\rm thr}$ of the physical thresholds with the scatter $\sigma_{\rm rot}$ in XUV fluence caused mainly by the unknown initial rotation of the hosts (Eq.~\ref{eq:sigmaF}). After accounting for $\sigma_{\rm rot}$, we obtain $\sigma_{\rm thr}=0.27\pm0.14$\,dex. If the contributions to this scatter are independent, the core-mass contribution cannot exceed the total, so that $a\,\sigma_{M_{\rm c}}\leq\sigma_{\rm thr}$. For planets near the centre $y_{\rm c}$ of the transition, this gives upper limits of $0.34$\,dex for $a=1.2$ and $0.20$\,dex for $a=2$, at $84\%$ probability (Appendix~\ref{app:mass}). Without the rotational scatter, these limits rise to $0.37$ and $0.22$\,dex.

The measured planet masses provide a complementary estimate of the mass dispersion at fixed $\mathcal{F}_Z$. Assuming that the envelopes contribute only a small fraction of the total mass, this dispersion approximates that of the cores. We obtain $\sigma_{M_p}=0.35\pm0.03$\,dex for the full measured-mass sample and $0.28\pm0.03$\,dex for masses measured at better than $3\sigma$ (Appendix~\ref{app:mass}). The former exceeds the limit for $a=2$ and is comparable to that for $a=1.2$, while the latter exceeds only the limit for $a=2$.

For comparison, the Rayleigh core-mass distribution adopted by \citet{OwenWu2017} to reproduce the radius valley has a dispersion of $0.28$\,dex, equal to the upper limit obtained for the slope of their model, $a=1.42$ (Appendix~\ref{app:mass}). If the measured planet-mass dispersions are representative of the core-mass dispersion near the transition, the relation $a\leq\sigma_{\rm thr}/\sigma_{M_{\rm c}}$ (Eq.~\ref{eq:sigmathr}) gives upper limits on $a$ of approximately $1.2$ and $1.4$ for the full and higher-precision samples, respectively, at $84\%$ probability.

These constraints favour a relatively shallow dependence of the threshold fluence on core mass, but remain conditional on the representativeness of the measured-mass sample, the approximation $M_p\simeq M_{\rm c}$, and the interpretation of the threshold width. More precise planet masses, particularly for planets near the transition, are needed to distinguish between a genuinely shallow mass dependence and limitations of the present mass measurements and population model.

%% file: sections/07a_discussion_mechanisms.tex
\section{Discussion}\label{sec:disc}

\subsection{Photoevaporation and core-powered mass loss}\label{sec:discmech}

Photoevaporation is driven by the XUV emission of the host, whereas core-powered mass loss is regulated by bolometric irradiation through the planetary equilibrium temperature \citep{GuptaSchlichting2019,GuptaSchlichting2020}. We therefore compare the threshold in accumulated XUV fluence $\mathcal{F}_{\rm XUV}$ with thresholds in two bolometric variables: the insolation $F$, which directly enters core-powered mass-loss models, and the bolometric fluence $\mathcal{F}_{\rm bol}$, used here as an empirical measure of the irradiation accumulated over the age of the system (Sect.~\ref{sec:axes}). At fixed $F$, the XUV fluence scales across the sample approximately as $M_\star^{-2.4}t^{0.4}$, whereas $\mathcal{F}_{\rm bol}$ scales as $M_\star^{-0.2}t^{1.0}$ (Appendix~\ref{app:host}).

The observed host-mass dependence is consistent with that encoded in $\mathcal{F}_{\rm XUV}$. At fixed insolation, a constant XUV threshold predicts a transition insolation increasing approximately as $F_{\rm thr}\propto M_\star^{2.4}$ at fixed age. The fitted threshold in $F$ gives $\xi_M=-3.26^{+1.05}_{-1.09}$, corresponding to $F_{\rm thr}\propto M_\star^{3.26}$, in agreement with this expectation. Its $95\%$ interval excludes both the much weaker mass dependence of $\mathcal{F}_{\rm bol}$ and the absence of an explicit mass dependence in a threshold based on $F$ alone (Sect.~\ref{sec:host}; Appendix~\ref{app:host}).

The age dependence provides a weaker indication in the same direction. When a residual term $\xi_t\log_{10}(t/{\rm Gyr})$ is added to the threshold, its coefficient is consistent with zero for $\mathcal{F}_{\rm XUV}$. For $F$ and $\mathcal{F}_{\rm bol}$, the coefficients differ from zero by approximately $1.5\sigma$, with opposite signs (Sect.~\ref{sec:age}). This suggests that insolation alone underpredicts the increase in SE weight with age, whereas bolometric fluence may overpredict it. Neither residual is statistically significant.

Within the Bondi-limited regime of core-powered mass loss, the observed dependence on host mass could be reproduced by a systematic increase in the core masses of planets near the atmospheric-loss threshold. The fitted threshold in insolation, including the host-mass term of Eq.~\eqref{eq:hostmass}, is
\begin{equation}\label{eq:fthr}
\log_{10}F_{\rm thr}=y_{\rm c}-\xi_M\log_{10}(M_\star/M_{\star,0})-\xi_Z\Delta,
\end{equation}
with $\xi_M=-3.26^{+1.05}_{-1.09}$. At fixed metallicity, the median threshold therefore grows with host mass as
\begin{equation}\label{eq:fthrmass}
F_{\rm thr}\propto M_\star^{-\xi_M}.
\end{equation}
For Bondi-limited core-powered escape, which applies at periods longer than about $8$\,d, the threshold insolation scales as
\begin{equation}\label{eq:corepowered}
F_{\rm thr}\propto\left(\frac{M_{\rm c}}{R_{\rm c}}\right)^{4}\propto M_{\rm c}^{2.9},
\end{equation}
where $R_{\rm c}\propto M_{\rm c}^{0.27}$ for a rocky core (Sect.~\ref{sec:massobs}). 
Matching the observed host-mass dependence would therefore require the characteristic core mass near the transition to scale approximately as $M_{\rm c}\propto M_\star^{-\xi_M/2.9}=M_\star^{1.1\pm0.4}$, assuming that other relevant planetary properties do not introduce a comparable dependence on host mass.

Photoevaporation does not require this additional core-mass scaling, because the reconstructed XUV fluence already contains a substantial dependence on host mass. Nevertheless, such a scaling is not excluded. In energy-limited escape, the threshold fluence scales as $M_{\rm c}^{a}$, with $a$ between $1.2$ and $2$ (Sect.~\ref{sec:massobs}). If $M_{\rm c}\propto M_\star^\nu$, the threshold in $\mathcal{F}_{\rm XUV}$ acquires a host-mass coefficient $\xi_M=-a\nu$. For $\nu=1.1$, the expected coefficient lies between $-1.3$ and $-2.2$, while the measured value is $\xi_M=-1.57^{+0.97}_{-0.86}$ (Sect.~\ref{sec:host}). The uncertainties therefore do not distinguish a constant characteristic core mass from a moderate increase with host mass.

The planetary radii provide a further consistency test. If the core masses of stripped planets increased as $M_\star^{1.1\pm0.4}$, their rocky-core radii would increase approximately as $M_\star^{0.30\pm0.10}$. 
Paper~II instead found a slope of the mean $\log_{10}R_p$ of the SE component with $\log_{10}M_\star$ of $-0.33^{+0.43}_{-0.40}$. With the reference threshold and the host-dependent mean radii of Eq.~\eqref{eq:meanradii}, the slope is $-0.14^{+0.39}_{-0.40}$ (Appendix~\ref{app:metal}). Neither measurement shows the expected increase, although the differences are only about $1.4\sigma$ and $1.1\sigma$, respectively.
The SN radii do increase with host mass, with a slope of $+0.35\pm0.16$, as reported in Paper~II. This may reflect a core-mass distribution extending to higher masses around more massive hosts without a corresponding increase in the mean mass of the stripped population.

Taken together, the dependence of the SE weight on host mass and its weaker dependence on age favour an interpretation based on cumulative XUV irradiation over the simple bolometric thresholds tested here. The absence of the expected increase in mean SE radius provides additional, but not decisive, support. These results are consistent with photoevaporation playing a major role in shaping the observed population, although they do not exclude core-powered mass loss or a contribution from both mechanisms. Moreover, a residual age dependence may remain below $1$\,Gyr, where Paper~II found a lower SE weight than predicted by the reference threshold, although the difference is not statistically significant (Sect.~\ref{sec:age}).

%% file: sections/07b_discussion_threshold.tex
\subsection{Interpretation of the threshold}\label{sec:discmass}

The reference threshold is characterised by its centre $y_{\rm c}$, its width $\sigma_{\mathcal{F}}$, and the background fraction $w_{\rm b}$ (Table~\ref{tab:params}), with the upper asymptote fixed at unity (Eq.~\ref{eq:W}). If the planetary envelopes are removed by energy-limited photoevaporation \citep{OwenWu2017,Mordasini2020}, $\mathcal{F}_{\rm c}=10^{y_{\rm c}}$ is the median XUV fluence $\mathcal{F}_{\rm XUV}$ at which the planets lose their envelopes, and the width $\sigma_{\mathcal{F}}$ limits the spread $\sigma_{M_{\rm c}}$ of their core masses (Sect.~\ref{sec:masswidth}). 
The two asymptotes describe the SE weight far below and above the transition, independently of the physical mechanism responsible for it.

For hosts at the median metallicity of $\mathcal{D}_3$, the threshold centre is $\mathcal{F}_{\rm c}=10^{21.13^{+0.10}_{-0.14}}$\,erg\,cm$^{-2}$. This can be compared with the XUV fluence required to remove an envelope if a fraction $\eta_{\rm esc}$ of the absorbed high-energy radiation is converted into work against its gravitational binding energy. For a rocky core of $M_{\rm c}=5\,M_\oplus$, an envelope of $M_{\rm env}=0.02\,M_{\rm c}$, and $\eta_{\rm esc}=0.1$, the estimated stripping fluence is approximately $10^{21.2}$\,erg\,cm$^{-2}$ (Appendix~\ref{app:massslopes}), comparable to $\mathcal{F}_{\rm c}$. Only order-of-magnitude agreement is expected, given the dependence on core and envelope masses, escape efficiency, and the characteristic radius at which XUV energy is absorbed.

The fluence-dependent transition does not explain the origin of the background SE fraction, whose fitted value is $w_{\rm b}=0.64^{+0.09}_{-0.10}$. This fraction refers to the intrinsic mixture population represented by the model, rather than only to detected planets, because the likelihood accounts for detection and confirmation probabilities (Sect.~\ref{sec:model}). In contrast, among detected planets with radii of $1$--$4\,R_\oplus$, the counted SE fraction $f_{\rm SE}^{\rm c}$ expected in the lowest-$\mathcal{F}_Z$ interval of Fig.~\ref{fig:ppc} is only $0.07\pm0.01$. At these fluences, planets generally have long orbital periods, where smaller SEs are less readily detected than SNs. Moreover, the adopted radius mixture places $40^{+6}_{-5}\%$ of the SE component below $1\,R_\oplus$, outside the range used to calculate $f_{\rm SE}^{\rm c}$. The importance of this model-dependent contribution is illustrated by restricting the fit to planets larger than $1\,R_\oplus$, which reduces $w_{\rm b}$ to $0.41\pm0.09$ (Appendix~\ref{app:forms}).

The remaining background fraction may include planets with stripping thresholds below the fluences sampled here, planets formed without substantial primordial envelopes \citep{OwenWu2017,OwenMurrayClay2018,NeilRogers2020}, or low-mass cores stripped during disc dispersal \citep{OwenWu2016,Rogers2024}. At the opposite end, the fitted upper asymptote leaves little room for a persistent SN component at high XUV fluence. When this asymptote is allowed to vary, its one-sided $95\%$ lower limit is $0.98$ (Appendix~\ref{app:forms}), consistent with nearly all planets belonging to the SE component at sufficiently high fluence within the adopted model.

Formation processes may also contribute to the radius valley, through late gas accretion \citep{LeeConnors2021} or the formation and inward migration of water-rich planets \citep{LuquePalle2022,Burn2024}. Such processes can establish differences in planetary radii and envelope properties before subsequent atmospheric evolution. However, Paper~II found that the SE weight increases with host age, and the present analysis shows that this increase can be described by the accumulation of XUV fluence without requiring an additional age dependence (Sect.~\ref{sec:age}). These results favour an evolutionary contribution to the observed radius valley, while leaving room for formation processes to determine part of the initial planet population, including a possible contribution to the background fraction $w_{\rm b}$.

%% file: sections/07c_discussion_metallicity_limits.tex
\subsection{Host metallicity}\label{sec:discfeh}

The convective turnover times, the luminosity evolution, and the rotational model of Paper~I are computed at solar metallicity,
although stellar high-energy emission and gyrochronological ages depend on metallicity 
\citep{Amard2020,AmardMatt2020}. 
The fitted coefficient $\xi_Z$ may therefore absorb metallicity-dependent differences in the XUV fluence and ages, as well as a genuine dependence of planetary properties or atmospheric loss on metallicity. These effects cannot be separated with the present data, particularly given the narrow metallicity range of $\mathcal{D}_3$, from $-0.22$ to $+0.24$ in [Fe/H] between the 5th and 95th percentiles (Sect.~\ref{sec:metal}). Nevertheless, the comparison of irradiation variables is insensitive to the inclusion of this term. With or without the metallicity term, the threshold in $\mathcal{F}_{\rm XUV}$ provides the highest log-likelihood, while those in $F$ and $\mathcal{F}_{\rm bol}$ perform less well (Table~\ref{tab:law}). Even without the metallicity term, the XUV threshold requires no additional dependence on host colour, and a host-mass term lowers its BIC by only $0.5$ (Sect.~\ref{sec:host}).

The evidence for a metallicity dependence is tentative (Sect.~\ref{sec:metal}). The negative coefficient $\xi_Z$ shifts the threshold towards higher $\mathcal{F}_{\rm XUV}$ around metal-rich hosts, consistent with envelopes being more resistant to removal. Both photoevaporation \citep{OwenMurrayClay2018} and core-powered mass loss \citep{GuptaSchlichting2020} can produce a dependence of this sign, so it does not distinguish between the mechanisms. The metallicity term improves the fit only at periods shorter than $10$\,d (Sect.~\ref{sec:metal}), where its amplitude can be compared with the occurrence rates measured by \citet{Petigura2018}. At periods of $1$--$10$\,d, they found that the occurrence rate of SNs increases with [Fe/H] more steeply than that of SEs, with the two populations separated at $1.7\,R_\oplus$. Their results imply a decrease in the logit of the SE fraction of $2.3\pm0.8$ per dex of [Fe/H]. For the planets of $\mathcal{D}_3$ in the same period range, we calculate the mean SE weights $\bar{w}_+$ and $\bar{w}_-$ predicted by the reference threshold after setting the metallicity offsets $\Delta_i$ of all hosts to $+0.2$ and $-0.2$, respectively, while keeping the other properties fixed. These offsets approximately span the metallicity range of the sample. The resulting slope, $[{\rm logit}\,\bar{w}_+-{\rm logit}\,\bar{w}_-]/0.4=-2.6^{+1.2}_{-1.5}$ per dex of [Fe/H], agrees with the occurrence-rate result of \citet{Petigura2018}, despite the different definitions of the SE and SN populations.

%% file: sections/08_conclusions.tex
\section{Conclusions}\label{sec:conclusions}

Using the \textit{Kepler} planets of Paper~II around FGK dwarfs, with gyrochronological ages from Paper~I and a likelihood that accounts for the detection and confirmation of each planet, we have investigated which irradiation variable best describes the intrinsic super-Earth fraction. Our principal findings are:

\begin{enumerate}

\item The accumulated XUV fluence $\mathcal{F}_{\rm XUV}$ describes the SE fraction better than insolation or bolometric fluence.

\item The preferred model describes the increase in SE weight through a threshold centred at $\mathcal{F}_{\rm c}=10^{21.13^{+0.10}_{-0.14}}\,{\rm erg\,cm^{-2}}$, with a width of $\sigma_{\mathcal{F}}=0.32^{+0.12}_{-0.10}$\,dex. The characteristic fluence is comparable, in order of magnitude, to that required to remove a modest envelope from a rocky core of a few Earth masses. The model also includes a background fraction $w_{\rm b}=0.64^{+0.09}_{-0.10}$, the SE weight below the transition, whose physical origin is not uniquely determined.

\item The reconstructed XUV fluence accounts for the observed dependence of the SE weight on host mass without requiring an additional mass term. In contrast, reproducing this dependence with the Bondi-limited scaling of core-powered mass loss would require characteristic core masses increasing approximately as $M_{\rm c}\propto M_\star^{1.1\pm0.4}$, a trend not evident in the measured SE radii. The accumulation of XUV fluence also describes the age dependence found in Paper~II without requiring an additional age term, although a residual below $1$\,Gyr remains possible. Together, these results favour a mass loss driven by the XUV irradiation over the simple bolometric thresholds tested here, without excluding a contribution from core-powered mass loss.

\item Under an energy-limited photoevaporation interpretation, the width of the threshold constrains the dispersion of core masses near the transition. If the stripping fluence scales as $M_{\rm c}^{a}$, the inferred dispersion in $\log_{10}M_{\rm c}$ is below $0.34$\,dex for $a=1.2$ and below $0.20$\,dex for $a=2$, the range expected from energy-limited escape \citep{OwenWu2017,Mordasini2020}, with $84\%$ probability. The available planet masses do not independently establish this scaling, because the observed mass--fluence relation is consistent with that induced by the mass--radius relation alone.

\item There is tentative evidence that the threshold shifts towards higher XUV fluence around metal-rich hosts. The inferred dependence is consistent with the metallicity trends in planet occurrence rates, but its physical origin remains uncertain because the reconstructed stellar activity histories and gyrochronological ages also depend on metallicity.

\end{enumerate}

The PLATO mission \citep{Rauer2024} is expected to provide precise radii and asteroseismic ages for many small-planet systems around bright stars, while ground-based radial-velocity follow-up will constrain their masses.
These observations will allow the XUV threshold to be tested with independently determined stellar ages and across a wider range of host properties.
In particular, improved planet masses and a broader metallicity range will help establish whether the threshold depends on core mass and composition as expected from atmospheric-loss models, and whether cumulative XUV irradiation is the principal variable governing the evolution of the radius valley.

%% file: sections/A_fluence_variables.tex
\section{The fluence variables}\label{app:axes}
The main-sequence luminosity $L_{\rm M}$ entering Eqs.~\eqref{eq:fbol} and \eqref{eq:fx} is interpolated linearly in stellar mass and logarithmic age from the solar metallicity MIST tracks \citep[version 1.2;][]{Dotter2016,Choi2016}. Only the main-sequence part of each track is used. Outside this interval, $L_{\rm M}$ is held at its first or last main-sequence values. Thus, the pre-main-sequence and post-main-sequence luminosity evolution is not explicitly modelled. 

The extreme-ultraviolet surface fluxes in the $10$--$36$ and $36$--$92$\,nm bands are obtained from the X-ray surface flux
\begin{equation}\label{eq:fxsurf}
f_{\rm X}=\frac{L_{\rm X}}{4\pi R_\star^2},
\end{equation}
in erg\,s$^{-1}$\,cm$^{-2}$.
We use the present radius $R_\star$ of each host throughout the calculation, rather than its time-dependent radius.
Following Eqs.~20 and 22 of \citet{Johnstone2021}:
\begin{equation}\label{eq:euv}
f_{\rm EUV,1}=110\,f_{\rm X}^{0.681},\qquad f_{\rm EUV,2}=0.924\,f_{\rm EUV,1}^{0.920}.
\end{equation}
The corresponding EUV luminosity that enters $R_{\rm XUV}$ (Sect.~\ref{sec:axes}) is
\begin{equation}\label{eq:leuv}
L_{\rm EUV}=4\pi R_\star^2\,\bigl(f_{\rm EUV,1}+f_{\rm EUV,2}\bigr).
\end{equation}
All fluences are calculated assuming that the planetary orbits remain unchanged over the age of their host.

\subsection{Rotational evolution and X-ray activity}
From $0.1$\,Gyr onwards, $R_{\rm X}(t')$ follows the rotation period $P_{\rm rot}(M_\star,t')$ that the model of Paper~I gives on the slow-rotator sequence, at solar metallicity, up to the end of the main sequence. The Rossby number is $Ro=P_{\rm rot}/\tau_{\rm c}$, where $\tau_{\rm c}(M_\star,t')$ is the convective turnover time of the stellar models of \citet{Spada2013}. $R_{\rm X}$ is then given by Eq.~17 of \citet{Johnstone2021},
\begin{equation}\label{eq:rxro}
R_{\rm X}=R_{\rm X,sat}\left(\frac{Ro}{Ro_{\rm sat}}\right)^{\beta},
\end{equation}
with $R_{\rm X,sat}=5.135\times10^{-4}$, $Ro_{\rm sat}=0.0605$, and $\beta=-0.135$ for $Ro\le Ro_{\rm sat}$ and $-1.889$ above. This relation is calibrated with the turnover times of \citet{Spada2013}.

Before $0.1$\,Gyr, where the model of Paper~I is not defined, we hold the rotation period at its value at $0.1$\,Gyr and let $\tau_{\rm c}$ decrease as the star contracts. $R_{\rm X}$ is thus continuous at $0.1$\,Gyr, and leaves the saturated regime between $14$\,Myr at $1.25\,M_\odot$ and $73$\,Myr at $0.60\,M_\odot$, and only after $0.1$\,Gyr below $0.55\,M_\odot$. The planets of $\mathcal{D}_3$ accumulate a median of $16\%$ of their $\mathcal{F}_{\rm XUV}$ before $0.1$\,Gyr. The rotation and activity tracks of \citet{Johnstone2021}, which follow the disc-locking phase and the pre-main-sequence contraction, and are calibrated on open clusters from $12$\,Myr onwards, give a similar $\mathcal{F}_{\rm XUV}$. With their track whose period at $0.1$\,Gyr equals that of Paper~I, the $\mathcal{F}_{\rm XUV}$ of each planet changes by at most $0.04$\,dex. With their median-rotation track, that of a star born with the median initial rotation rate, the $\mathcal{F}_{\rm XUV}$ of each planet changes by between $-0.03$ and $+0.08$\,dex.

Rotation periods and turnover times are evaluated at masses from $0.50$ to $1.25\,M_\odot$ in steps of $0.05\,M_\odot$, which cover the hosts of $\mathcal{D}_3$ ($0.54$ to $1.23\,M_\odot$). We take the turnover times of the models of \citet{Spada2013} as tabulated along the tracks of \citet{Johnstone2021}, which are published up to $1.2\,M_\odot$. For the two hosts slightly above this mass ($1.21$ and $1.23\,M_\odot$, with three planets in total), we compute a track at $1.25\,M_\odot$ with the same model. We interpolate $\log_{10}R_{\rm X}$ linearly in mass and in $\log_{10}t'$, and hold it at its value at $1$\,Myr at earlier ages and at its last value beyond the end of the main sequence.

The initial rotation of each host is unknown. When the initial rotation rate of each host is drawn, independently for each age draw, from the distribution of \citet{Johnstone2021}, and their tracks are used at all ages, $\mathcal{F}_{\rm XUV}$ has a standard deviation of $\sigma_{\rm rot}=0.17$\,dex among hosts of the same mass. This scatter is not included in $\mathcal{F}_{\rm XUV}$ and contributes to the width $\sigma_{\mathcal{F}}$ of the threshold (Sect.~\ref{sec:masswidth}). Such a scatter is expected to broaden the transition rather than to shift it: in the simulations of \citet{KetzerPoppenhaeger2023}, a spread of activity evolution among stars of the same mass makes the lower edge of the radius valley less sharp without significantly changing its location.

Table~\ref{tab:rotvariants} gives the fit of the reference threshold with other choices of $R_{\rm X}$ before $0.1$\,Gyr, with Eq.~\eqref{eq:rxro} unchanged at later ages: the median-rotation track, the track of \citet{Johnstone2021} whose period at $0.1$\,Gyr equals that of Paper~I, and a saturated activity, $R_{\rm X}=R_{\rm X,sat}$. It also gives two choices that change $R_{\rm X}$ after $0.1$\,Gyr as well: the median-rotation track kept until it reaches the slow-rotator sequence, for hosts up to $0.95\,M_\odot$, and the tracks of \citet{Johnstone2021} at all ages. The reference choice requires no rotation or activity model beyond those used after $0.1$\,Gyr, and its $\mathcal{F}_{\rm XUV}$ is consistent with that of the other choices: the median difference is $0.01$\,dex or less with the tracks of \citet{Johnstone2021} before $0.1$\,Gyr, and at most $0.05$\,dex in the other cases. Across these choices, the log-likelihood of the reference threshold changes by at most $0.4$, and exceeds that of the threshold in $P$ by $2.1$ to $2.8$.
\begin{table}
\caption{Fit of the reference threshold with other choices of $R_{\rm X}$ (Appendix~\ref{app:axes}). Tracks are the rotation and activity tracks of \citet{Johnstone2021}; the median-rotation track is that of a star born with the median initial rotation rate. $\Delta\log_{10}\mathcal{F}_{\rm XUV}$: change in fluence, in dex (median and 16th--84th percentiles over planets and age draws). $\Delta\ln\mathcal{L}$: change in the log-likelihood of the reference threshold at the MAP. $\Delta\ln\mathcal{L}_P$: its difference from the threshold in $P$.}
\label{tab:rotvariants}
\centering\small\setlength{\tabcolsep}{3.5pt}\renewcommand{\arraystretch}{1.35}
\begin{tabular}{@{}lccc@{}}
\hline\hline
Choice of $R_{\rm X}$ & $\Delta\log_{10}\mathcal{F}_{\rm XUV}$ & $\Delta\ln\mathcal{L}$ & $\Delta\ln\mathcal{L}_P$\\
\hline
\noalign{\smallskip}
\multicolumn{4}{@{}l}{\textit{Before $0.1$\,Gyr}}\\
\noalign{\smallskip}
Constant period (reference) & $0$ & $0$ & $+2.47$\\
Median-rotation track & $+0.011^{+0.016}_{-0.011}$ & $+0.28$ & $+2.75$\\
Track matching Paper~I at $0.1$\,Gyr & $-0.004^{+0.004}_{-0.005}$ & $-0.11$ & $+2.36$\\
Saturated, $R_{\rm X}=R_{\rm X,sat}$ & $+0.048^{+0.025}_{-0.020}$ & $-0.39$ & $+2.08$\\
\noalign{\smallskip}
\hline
\noalign{\smallskip}
\multicolumn{4}{@{}l}{\textit{Median-rotation track, also after $0.1$\,Gyr}}\\
\noalign{\smallskip}
Up to the slow-rotator sequence & $+0.026^{+0.064}_{-0.026}$ & $+0.37$ & $+2.84$\\
At all ages & $-0.030^{+0.052}_{-0.016}$ & $+0.21$ & $+2.68$\\
\hline
\end{tabular}
\end{table}

The fit is insensitive to two further choices. First, we multiply $L_{\rm X}$ by $10^{0.359\,g}$, with $g$ drawn from a standard normal distribution independently for each planet and age draw, where $0.359$\,dex is the scatter about Eq.~\eqref{eq:rxro} \citep{Johnstone2021}. The log-likelihood of the reference threshold decreases by $1.1$. This overestimates the effect, since the scatter arises mostly from the variability of each star on timescales of decades \citep{Johnstone2021}, which averages out in the time integral. Second, with $\mathcal{F}_{\rm X}$ in place of $\mathcal{F}_{\rm XUV}$, the log-likelihood increases by $0.1$, so that the data do not distinguish the X-ray from the XUV fluence.

\subsection{Cluster calibration}
The XUV fluence $\mathcal{F}^{\rm J}_{\rm XUV}$ (Sect.~\ref{sec:axes}) replaces the rotational evolution above with the mean activity evolution of open clusters, to test whether the comparison of the irradiation variables depends on the activity model. In the open-cluster calibration of \citet{Jackson2012}, the ratio $R_{\rm X}$ of Eq.~\eqref{eq:fx} is
\begin{equation}\label{eq:rx}
R_{\rm X}(t)=\begin{cases}R_{\rm sat}, & t\le\tau_{\rm sat},\\ R_{\rm sat}\,(t/\tau_{\rm sat})^{-\alpha}, & t>\tau_{\rm sat},\end{cases}
\end{equation}
where the saturated value $R_{\rm sat}$, the age $\tau_{\rm sat}$ at which the decline begins, and the slope $\alpha$ of the decline are taken from their Table~2 for each of their seven bins of de-reddened colour $(B-V)_0$, which span $0.29$ to $1.41$.
We interpolate $\log_{10}R_{\rm sat}$, $\log_{10}\tau_{\rm sat}$, and $\alpha$ linearly in $(B-V)_0$ between the centres of these bins, and keep the values of the first and last centre for bluer and redder hosts.
We convert the de-reddened \textit{Gaia} colour $(G_{\rm BP}-G_{\rm RP})_0$ into $(B-V)_0$ along the dwarf sequence of \citet{PecautMamajek2013}, version 2022.04.16 of their online table\footnote{\url{http://www.pas.rochester.edu/~emamajek/EEM_dwarf_UBVIJHK_colors_Teff.txt}}. 
The calibration of \citet{Jackson2012} constrains the decline up to the age of Praesepe, $\sim$$0.7$\,Gyr, and is extrapolated beyond it. In three of its colour bins, few clusters lie in the unsaturated regime. 
These uncertainties do not change the comparison of the irradiation variables. In $62$ realisations, we drew $R_{\rm sat}$, $\tau_{\rm sat}$, and $\alpha$ of each bin independently from their fitting uncertainties in \citet{Jackson2012}, and refitted the threshold in $\mathcal{F}^{\rm J}_{\rm XUV}$ without the metallicity term. Its log-likelihood changes by a median of $-0.44$, with 16th--84th percentiles of $-1.33$ and $+0.23$. In two of these realisations the threshold in $\mathcal{F}^{\rm J}_{\rm XUV}$ fits the data better than the threshold in $P$, and in none better than the threshold in $\mathcal{F}_{\rm XUV}$.

%% file: sections/B_model_details.tex
\section{Details of the model}\label{app:model}

\subsection{Fits}\label{app:fits}
The MAP of each form is found by Nelder--Mead minimisation \citep{NelderMead1965} from several starting points. The chains of the threshold forms have $40$ walkers and between $4000$ and $10000$ steps after the burn-in, which is the first third of the initial run. For every parameter, the potential scale reduction factor $\hat R$ of \citet{GelmanRubin1992}, computed on the two halves of the chain of each walker, is at most $1.04$ in the chain of the reference threshold and $1.05$ in the others. The BIC counts the $581$ planets as data. Since the planets of a host share its age, mass, and metallicity, the effective number of independent data lies between the $404$ hosts and the $581$ planets. With $404$, the ranking of the threshold forms by BIC does not change.

The confirmation probability of Paper~II (Sect.~\ref{sec:model}) varies with host age through a term $\mathcal{H}$. It is fixed at zero in the MAP fits and free in the chains, with its profile likelihood from the DR25 candidates as prior, $\mathcal{H}=+0.033^{+0.036}_{-0.038}$\,dex\,Gyr$^{1/2}$ (Paper~II). The chains that compare the two colour subsets (Sects.~\ref{sec:linear} and~\ref{sec:host}) hold it at the median of this prior, as in Paper~II. Since this prior is close to zero, $\mathcal{H}$ has little effect on the fits. With $\mathcal{H}$ at its median instead of zero, the log-likelihoods of the threshold forms at the MAP change by at most $0.53$, and their ranking by BIC is unchanged. In the chain of the reference threshold, the posterior of $\mathcal{H}$ is $+0.020^{+0.034}_{-0.035}$\,dex\,Gyr$^{1/2}$.

The mixture parameters have the bounds of Paper~II, and all the other priors are flat. For the threshold forms, $w_{\rm b}$ lies in $[0,0.95]$, $\sigma_y$ in $[0.02,3]$ in the units of the irradiation variable $h$, and the coefficients $\xi$ in $[-5,5]$. The prior of $y_{\rm c}$ spans the range that contains the central $99\%$ of the values of $h$ over $\mathcal{D}_3$, plus half of its width on each side. In the chain of the reference threshold no posterior is limited by these bounds. The posterior of $\xi_M$ (Eq.~\eqref{eq:hostmass}) in the threshold in $F$ extends below $-5$. With the metallicity term, a chain with $\xi_M$ in $[-10,10]$ and $\mathcal{H}$ fixed at zero gives $\xi_M=-3.16^{+1.20}_{-1.18}$, with a $95\%$ interval from $-5.52$ to $-0.65$, close to the profile likelihood (Appendix~\ref{app:host}). The profile likelihood of a coefficient (Sect.~\ref{sec:fits}) is computed on a grid of its values, with the other parameters refitted at each point. Since it does not depend on the prior, the grid of $\xi_M$ extends below $-5$. The $68\%$ and $95\%$ intervals contain the values of the coefficient at which the profile likelihood lies below its maximum by $\Delta\ln\mathcal{L}<0.5$ and $1.92$, respectively. The intervals of $f_{\rm SE}^{\rm c}$ are from a bootstrap over hosts.

\subsection{Background fraction and shape of the transition threshold}\label{app:forms}
Equation~\eqref{eq:W} has two features that we test here: the background fraction $w_{\rm b}$ and the normal shape of the transition threshold between the SE and SN components.
With $w_{\rm b}=0$ and a logistic transition, the SE weight rises from zero to unity in $\log_{10}\mathcal{F}_Z$. Its log-likelihood is $9.5$ below that of the reference threshold and $1.3$ below that of the model of Paper~II (Eq.~\eqref{eq:wpaperii}).
The value of $w_{\rm b}$ depends on the shape assumed for the SE component, which extends below $1\,R_\oplus$, where few planets are detected. Considering only the $541$ planets with $R_p\ge1\,R_\oplus$, the reference threshold has $w_{\rm b}=0.41\pm0.09$, compared with $0.66$ at the MAP for all the planets. With a skew-normal SE component, $w_{\rm b}=0.55$. In both cases each $\Delta\ln\mathcal{L}$ of Table~\ref{tab:evidence} between threshold forms keeps its sign. Considering only the planets with $R_p\ge1\,R_\oplus$, the gain of the metallicity term falls from $4.3$ to $2.9$, and the $\Delta\ln\mathcal{L}$ of the reference threshold over the threshold in $P$ from $2.5$ to $1.0$. With the skew-normal component, the gain of the metallicity term falls to $2.7$.

None of the other shapes of the transition threshold that we tested improves the fit significantly. At the same number of parameters, the log-likelihood changes by $-0.4$ to $+0.4$ when $\Phi$ is replaced with the cumulative distribution function of the logistic or of either Gumbel distribution, or when $\mathcal{F}_{\rm XUV}$ is computed with the other choices of $R_{\rm X}$ of Table~\ref{tab:rotvariants}. With $\Phi$ replaced by the cumulative distribution function of a split normal, with mode at $y_{\rm c}$, width $\sigma_{\mathcal{F}}$ below it and $r\,\sigma_{\mathcal{F}}$ above it, and $r$ free, the log-likelihood increases by $0.8$, which is less than the $3.18$ above which the BIC favours a form with one more parameter (Table~\ref{tab:evidence}), so that the split normal is not preferred. The maximum lies at $r=0$, where $\Theta$ reaches unity at $y_{\rm c}$.
The upper asymptote of $\Theta$, unity in Eq.~\eqref{eq:W}, remains unity at the MAP when it is left free, with a one-sided $95\%$ lower limit of $0.98$. At most $2\%$ of the planets at the highest irradiation thus belong to the SN component.

\subsection{The metallicity term}\label{app:metal}
The corrected $p$-value of Sect.~\ref{sec:metal} is the $p$-value of the metallicity term multiplied by five, the number of terms tested on the same data: the terms in the colour subset, in $P$, and in host mass added to the argument of $\Theta$, the metallicity term, and the age shifts $\delta_m$ (Sect.~\ref{sec:age}). The irradiation variables, choices of $R_{\rm X}$, and shapes of the transition threshold compared on the same data add further tests, so that the corrected $p$-value is a lower bound.

In the variant of the reference threshold of Sect.~\ref{sec:metal} with host-dependent mean radii, the mean $\log_{10}R_p$ of component ${\rm k}\in\{{\rm SE},{\rm SN}\}$ is
\begin{equation}\label{eq:meanradii}
\overline{\log_{10}R_p}_{,{\rm k}}=\mu_{\rm k}+\kappa^{M}_{\rm k}\log_{10}\frac{M_\star}{M_{\star,0}}+\kappa^{\rm Fe}_{\rm k}\Delta_i,
\end{equation}
where $\mu_{\rm k}$ is the mean at the median host mass $M_{\star,0}$ (Sect.~\ref{sec:host}) and at $\Delta_i=0$, and $\kappa^{M}_{\rm k}$ and $\kappa^{\rm Fe}_{\rm k}$ are its slopes in host mass and metallicity. The four coefficients increase the log-likelihood of the reference threshold by $16.3$ ($\Delta{\rm BIC}=-7.2$). In this variant, the metallicity term increases the log-likelihood by $2.0$ instead of $4.3$, and no longer lowers the BIC: without it, the variant in $\mathcal{F}_{\rm XUV}$ has a BIC lower by $2.4$. With host-dependent mean radii in all the threshold forms, $\mathcal{F}_{\rm XUV}$ keeps the highest log-likelihood of the five variables, with or without the metallicity term, by at least $1.4$. A chain of this variant, with $\mathcal{H}$ fixed at zero, gives $\kappa^{M}_{\rm SE}=-0.14^{+0.39}_{-0.40}$, so that the mean radius of the SE component does not grow with host mass.

The fits of this work use the measured [Fe/H] of each host. To propagate the uncertainties of the metallicities, we replace each measured value with $200$ draws from its posterior. The prior is the [Fe/H] distribution of the sample, corrected for the uncertainties and described either by the sum of two normal distributions, preferred by the BIC ($\Delta{\rm BIC}=30.5$), or by a single one. With the first prior, the posterior median of $\xi_Z$ in the reference threshold is $-1.93\pm0.77$, compared with $-1.17^{+0.40}_{-0.46}$ with the measured [Fe/H]. With the second, $\xi_Z=-1.34$ at the MAP, so that the correction depends on the prior. With the first prior, $\sigma_{\mathcal{F}}=0.33^{+0.15}_{-0.12}$\,dex, as with the measured [Fe/H], and the $\Delta\ln\mathcal{L}$ of $\mathcal{F}_{\rm XUV}$ over $P$ is $2.2$, compared with $2.5$.

\subsection{Host colour and host mass}\label{app:host}
Table~\ref{tab:variants} lists the thresholds of Sect.~\ref{sec:host} with a term in host mass, colour subset, or orbital period added to the argument of $\Theta$. At fixed $F$, each of the other irradiation variables $h$ also depends on host mass and age. We compare this dependence with the host-mass slope $\xi_M$ of the threshold in $F$. For the period we take $h=-\frac{4}{3}\log_{10}P$, by Kepler's third law. For each $h$, we regress $h-\log_{10}F$ on $\log_{10}M_\star$ and $\log_{10}t$ over all the age draws. The regression coefficients $\hat\xi_M$ and $\hat\xi_t$ give
\begin{equation}\label{eq:implied}
h\simeq\log_{10}F+\hat\xi_M\log_{10}\frac{M_\star}{M_{\star,0}}+\hat\xi_t\log_{10}\frac{t}{\rm Gyr}.
\end{equation}
The coefficients $\hat\xi_M$ and $\hat\xi_t$ use only the irradiation variables, host masses, and ages of the planets of $\mathcal{D}_3$, and not their radii, whereas $\xi_M$ is fitted to the radii with the likelihood of Sect.~\ref{sec:model}. The pairs $(\hat\xi_M,\hat\xi_t)$ are $(-4.24,-0.04)$ for $P$, $(-3.75,+0.20)$ for $\mathcal{F}^{\rm J}_{\rm XUV}$, $(-2.42,+0.42)$ for $\mathcal{F}_{\rm XUV}$, and $(-0.19,+0.96)$ for $\mathcal{F}_{\rm bol}$.

\begin{table}
\caption{Threshold forms with a term added to the argument of $\Theta$, without and with the metallicity term. The columns `without' and `with' give the $\Delta\ln\mathcal{L}$ of the forms without and with the metallicity term, at the MAP and relative to the reference threshold, as in Table~\ref{tab:law}. The last row is the reference threshold of Table~\ref{tab:law}, without and with the metallicity term. The other forms have nine parameters without the metallicity term and ten with it, so that $\Delta{\rm BIC}=-2\Delta\ln\mathcal{L}$ and $6.36-2\Delta\ln\mathcal{L}$, respectively. $\xi_Z$ is as in Table~\ref{tab:law}.}
\label{tab:variants}
\centering\small\setlength{\tabcolsep}{4pt}
\begin{tabular}{lccc}
\hline\hline
Threshold variable & without & with & $\xi_Z$\\
\hline
$\log_{10}\mathcal{F}_{\rm XUV}$ and $\log_{10}M_\star$ & $-0.8$ & $+1.2$ & $-0.68$\\
$\log_{10}\mathcal{F}_{\rm XUV}$ and colour subset & $-2.3$ & $+1.3$ & $-0.85$\\
$\log_{10}\mathcal{F}_{\rm XUV}$ and $\log_{10}P$ & $-3.1$ & $+0.8$ & $-1.39$\\
$\log_{10}F$ and $\log_{10}M_\star$ & $-2.5$ & $-0.4$ & $-0.82$\\
\hline
$\log_{10}\mathcal{F}_{\rm XUV}$ (reference) & $-4.3$ & $0$ & $-1.09$\\
\hline
\end{tabular}
\end{table}

The $\xi_M=-3.26^{+1.05}_{-1.09}$ of the threshold in $F$ with the metallicity term (Sect.~\ref{sec:host}) has a $95\%$ interval from $-5.40$ to $-1.17$. Its $68\%$ interval contains the $\hat\xi_M$ of $P$, $\mathcal{F}^{\rm J}_{\rm XUV}$, and $\mathcal{F}_{\rm XUV}$, and that of $\mathcal{F}_{\rm bol}$ lies outside the $95\%$ interval. Without the metallicity term, $\xi_M=-4.53^{+0.99}_{-0.95}$, and the $\hat\xi_M$ of $\mathcal{F}_{\rm XUV}$ lies just outside its $95\%$ interval, which extends from $-6.3$ to $-2.6$. The dependence on host mass that the threshold in $F$ requires is thus the one contained in $\mathcal{F}_{\rm XUV}$, $\mathcal{F}^{\rm J}_{\rm XUV}$, and $P$, and not the much weaker one of $\mathcal{F}_{\rm bol}$.

\subsection{Tests of the reference threshold}\label{app:tests}
Table~\ref{tab:evidence} gives three tests of the preferences that Sects.~\ref{sec:law} and~\ref{sec:age} find between models of the SE weight: on planets held out of the fit, on simulated catalogues (Appendix~\ref{app:mocks}), and on a bootstrap over hosts.

\begin{table}
\caption{Tests of the preferences that Sects.~\ref{sec:law} and~\ref{sec:age} find between models of the SE weight. Each row compares two models, listed as first vs second. The columns `all' and `held-out' give the $\Delta\ln\mathcal{L}$ between them on all planets and on planets held out of the fit, and a positive $\Delta\ln\mathcal{L}$ favours the first.}
\label{tab:evidence}
\centering\small\setlength{\tabcolsep}{3.5pt}
\begin{tabular}{@{}lccccc@{}}
\hline\hline
First vs second model & $\Delta n_{\rm par}$ & all & held-out & $N_{\rm sim}$ & $N_{\rm boot}$\\
\hline
\noalign{\smallskip}
\multicolumn{6}{@{}l}{\textit{Model of the weight}}\\
\noalign{\smallskip}
Reference vs Paper~II & $0$ & $+8.2$ & $+6.7$ & $0/200$ & $185/200$\\
\noalign{\smallskip}
\hline
\noalign{\smallskip}
\multicolumn{6}{@{}l}{\textit{Threshold variable}}\\
\noalign{\smallskip}
$\mathcal{F}_{\rm XUV}$ vs $F$ & $0$ & $+4.6$ & $+8.6$ & $1/200$ & $179/200$\\
$\mathcal{F}_{\rm XUV}$ vs $\mathcal{F}_{\rm bol}$ & $0$ & $+4.1$ & $+7.4$ & $1/200$ & $168/200$\\
$\mathcal{F}_{\rm XUV}$ vs $P$ & $0$ & $+2.5$ & $+0.5$ & $3/500$\tablefootmark{a} & $157/200$\\
\noalign{\smallskip}
\hline
\noalign{\smallskip}
\multicolumn{6}{@{}l}{\textit{With vs without}}\\
\noalign{\smallskip}
metallicity term & $1$ & $+4.3$ & $-0.3$ & $3/200$ & $146/200$\\
age shifts $\delta_m$ & $2$ & $+2.7$ & $+0.2$ & $19/200$ & $33/200$\\
\hline
\end{tabular}
\tablefoot{Reference: the reference threshold. In the comparisons of variables, both threshold models have the metallicity term. In the last two rows, the first model has the named term and the second does not: the metallicity term is tested in the threshold in $\mathcal{F}_{\rm XUV}$, and the age shifts in the reference threshold. All: the $581$ planets, at the MAP. Held-out: the $377$ planets of K and late-G hosts, for fits to the $204$ planets of F and G hosts. $N_{\rm sim}$: of the catalogues simulated from the second model, the number in which the first model gains at least the observed $\Delta\ln\mathcal{L}$ on all planets. $N_{\rm boot}$: the $404$ hosts are resampled $200$ times with replacement, each with all its planets, and $N_{\rm boot}$ is the number of resamplings in which the BIC favours the first model. $\Delta n_{\rm par}$: number of parameters of the first model in excess of the second. With $N=581$, the BIC favours the first model when $\Delta\ln\mathcal{L}$ exceeds $\Delta n_{\rm par}\ln N/2$, that is $0$, $3.18$, and $6.36$ for $\Delta n_{\rm par}=0$, $1$, and $2$, respectively. \tablefoottext{a}{$36/500$ for catalogues simulated from the threshold in $P$ that also includes the age shifts $\delta_m$, fitted with the same two models (Appendix~\ref{app:mocks}).}}
\end{table}

In a held-out test, the models are fitted to a part of the planets, and their log-likelihoods are compared on the other planets, which do not enter the fit. In the test of Table~\ref{tab:evidence}, the models are fitted to the $204$ planets of F and G hosts and evaluated on the $377$ planets of K and late-G hosts. A second held-out test fits the models to the planets younger than $2$\,Gyr and evaluates them on the $394$ planets at $2$--$4$\,Gyr. In both held-out tests, the thresholds in $F$ and in $\mathcal{F}_{\rm bol}$ have a lower log-likelihood on the held-out planets than that of the reference threshold. The difference ranges from $2.5$ to $13.0$ over the two tests, the two variables, and the forms with and without the metallicity term. The threshold in $\mathcal{F}^{\rm J}_{\rm XUV}$ instead has a higher log-likelihood than that in $\mathcal{F}_{\rm XUV}$, by $1.1$ to $6.7$. In the second test, the metallicity term raises the held-out log-likelihood by $1.9$. The held-out planets thus confirm the preference for $\mathcal{F}_{\rm XUV}$ over $F$ and $\mathcal{F}_{\rm bol}$, but not that for $\mathcal{F}_{\rm XUV}$ over $\mathcal{F}^{\rm J}_{\rm XUV}$, and they support the gain of the metallicity term only in the second test.

The standard deviations with which Sect.~\ref{sec:ppc} compares the counted and the expected fractions are those of their difference in each interval. They are computed with a bootstrap over hosts, in which the expected fraction also varies over the posterior of the reference threshold. To measure the agreement between the counted and the expected fractions over a set of intervals $b$, we use the sum of their squared differences, each in units of its variance,
\begin{equation}\label{eq:discrepancy}
T=\sum_b\frac{\bigl(f^{\rm c}_{{\rm SE},b}-E_b\bigr)^2}{V_b},
\end{equation}
where $E_b$ is the expected fraction of Sect.~\ref{sec:ppc} and $V_b$ the variance of $f^{\rm c}_{{\rm SE},b}$ for independent planets with different probabilities of being counted as SEs. Catalogues simulated from the posterior (Appendix~\ref{app:mocks}) give a $T$ larger than the observed one in $65\%$ of cases over the seven intervals of $\log_{10}\mathcal{F}_Z$ of Fig.~\ref{fig:ppc}, and in $56\%$ over the fourteen intervals of the two colour subsets. The counted fractions are thus consistent with the reference threshold, also in the two colour subsets separately.

The reference threshold assumes that the SE and SN components of the radius distribution have the same means and widths at every $\mathcal{F}_Z$. Section~\ref{sec:ppc} tests this assumption on five intervals of $\log_{10}\mathcal{F}_Z$ of equal population, with the $w_i$ of the reference threshold. In a first test, the mean $\log_{10}R_p$ of each component is fitted separately in each interval, and a $\chi^2$ test compares the five means with the mean common to all planets. In a second test, means and widths are both free in each interval, and a likelihood-ratio test compares this fit with the one with common values. Neither test detects a change.

For the SE component, however, the first test might not detect a shift of its mean $\log_{10}R_p$ along the valley locus. Such a shift is expected if more massive cores are stripped at higher XUV fluence. The valley locus of Eq.~\eqref{eq:locus} lies at larger radii at shorter periods. The median period is $34$\,d in the interval of lowest $\mathcal{F}_Z$, with $\log_{10}(\mathcal{F}_Z/{\rm erg\,cm^{-2}})$ from $19.2$ to $20.4$, and $2.2$\,d in that of highest $\mathcal{F}_Z$, from $21.4$ to $22.6$. Between these two median periods, the locus rises by $0.11$\,dex in $\log_{10}R_p$. If the mean $\log_{10}R_p$ of the SE component in the five intervals followed the locus, the first test would give $p=0.13$, compared with the observed $p=0.47$ (Sect.~\ref{sec:ppc}), so that it would not reject common means in either case. The fitted means of the SE component decrease with the median period of the interval with a slope of $-0.07\pm0.04$, compatible with that of the locus, $-0.096$, and at $2\sigma$ from zero. For the SN component, the slope is $+0.02\pm0.03$, compatible with zero and at $3.5\sigma$ from that of the locus. Adding a term $\kappa^{P}_{\rm k}\log_{10}(P/P_{\rm ref})$ to the mean radii of Eq.~\eqref{eq:meanradii}, one for each component, increases the log-likelihood by $2.7$, which the BIC does not favour for two extra parameters, and the $\Delta\ln\mathcal{L}$ of $\mathcal{F}_{\rm XUV}$ over $P$ is $3.5$, so that the comparison of the variables does not depend on this assumption.

\subsection{The age dependence}\label{app:age}
In Sect.~\ref{sec:age}, $\xi_t$ differs from zero by $1.6\sigma$ and $1.5\sigma$ for the thresholds in $F$ and in $\mathcal{F}_{\rm bol}$, respectively. Fixing $\xi_t=0$ lowers the log-likelihood by $\Delta\ln\mathcal{L}=1.3$ and $1.1$, respectively, and the significances are $(2\Delta\ln\mathcal{L})^{1/2}$. Neither coefficient thus differs significantly from zero.

Below $1$\,Gyr, $f_{\rm SE}^{\rm c}$ lies below the fraction expected from the reference threshold (Sect.~\ref{sec:age}). For the $115$ planets of $1$--$4\,R_\oplus$ at $1$--$2$\,Gyr and the $371$ at $2$--$4$\,Gyr, it lies above it, by $0.09\pm0.04$ and $0.02\pm0.02$. The deficit and the excess at $1$--$2$\,Gyr do not come from the selection: among the $253$ planets whose selection probability exceeds $0.8$ for a true radius of $1.25\,R_\oplus$, they are $-0.10^{+0.08}_{-0.09}$ and $+0.12^{+0.07}_{-0.06}$. Each of these differences lies at most about $2.3$ standard deviations from zero, so that none is significant on its own.

With $\mathcal{F}^{\rm J}_{\rm XUV}$ in place of $\mathcal{F}_{\rm XUV}$, the shifts $\delta_m$ remain positive and give $\Delta\ln\mathcal{L}=3.5$. For the reference threshold fitted separately to the two colour subsets, with $\xi_Z$ fixed at its MAP, the shifts are $\delta_{1\text{--}2}=+1.83^{+0.78}_{-0.70}$ and $\delta_{2\text{--}4}=+1.17^{+0.70}_{-0.62}$ for the K and late-G hosts, and $-0.6^{+1.5}_{-1.7}$ and $-0.4^{+1.2}_{-1.3}$ for the F and G hosts. Imposing equal shifts in the two subsets lowers the log-likelihood by $1.0$ ($p=0.36$), and a $w_{\rm b}$ that differs between them, with no dependence on age, gives only $\Delta\ln\mathcal{L}=0.7$. The residual is thus not specific to the adopted activity evolution and is measured only in the K and late-G hosts, although the two subsets do not differ significantly, and a difference in $w_{\rm b}$ between them does not account for it.

%% file: sections/C_simulated_catalogues.tex
\section{Simulated catalogues}\label{app:mocks}
We use simulated catalogues for two purposes. First, they show how often a difference in log-likelihood between two models of the SE weight, as large as the observed one, arises by chance (Table~\ref{tab:evidence}). Second, they give the counted SE fractions and the radius distributions expected from the reference threshold (Sect.~\ref{sec:ppc}).

Each catalogue is simulated from one model of the SE weight, with its parameters at the MAP or taken from a sample of the posterior. The models are the reference threshold and the threshold in $P$ with the metallicity term, the latter also with the age shifts $\delta_m$ of Sect.~\ref{sec:age}. Each catalogue keeps the hosts, orbits, and age draws of the $581$ planets of $\mathcal{D}_3$, and the probability of detection and confirmation of each planet as a function of its radius. For each planet, we choose one of its age draws at random, draw its component, SE or SN, with the SE weight of the model at that age, and draw its true radius from that component. The simulated planet is then detected and confirmed with the probability of Paper~II at that radius, computed with the age term $\mathcal{H}$ of the confirmation probability set to zero (Appendix~\ref{app:fits}). If it is not detected and confirmed, we draw a new component and a new radius with the same age draw, until the simulated planet is detected and confirmed, so that each catalogue contains one planet for each planet of $\mathcal{D}_3$. The measured radius then scatters around the true one according to its uncertainty and must fall within the radius range of the sample. When the reference threshold is refitted to catalogues simulated from it, the mean of each fitted parameter differs from the true value by at most $0.20$ times the standard deviation of that parameter across the catalogues.

We use four sets of catalogues. The first three are refitted at the MAP and test the comparisons of Table~\ref{tab:evidence}. The fourth gives the expected distributions of Sect.~\ref{sec:ppc}.
The first set tests the comparison of the XUV fluence and the period. We draw $500$ simulated catalogues from each of the two thresholds with the metallicity term, at their MAP. For each catalogue, we fit both thresholds to it and compute the difference of their log-likelihoods, $\ln\mathcal{L}_{\mathcal{F}_{\rm XUV}}-\ln\mathcal{L}_{P}$, which is $2.47$ for the data. This difference is at least $2.47$ in $3$ of the $500$ catalogues drawn from the threshold in $P$ and in $406$ of the $500$ drawn from the threshold in $\mathcal{F}_{\rm XUV}$. In $100$ further simulated catalogues from each threshold, each host is simulated at one age drawn from its uncertainty, and the fit uses new age draws, independent of that age. The two counts are then $1$ and $79$. A difference as large as the observed one is thus rare, about $1\%$, if the SE weight follows the threshold in $P$, and typical if it follows the threshold in $\mathcal{F}_{\rm XUV}$.

The second set tests whether a threshold in $P$ with an age dependence reproduces the observed difference. Its $500$ simulated catalogues are drawn from the threshold in $P$ with the metallicity term and with the age shifts $\delta_m$ of Eq.~\eqref{eq:ageshift}, which shift the logit of the SE weight of each age draw according to its age interval, with $\delta_{1\text{--}2}=+1.82$ and $\delta_{2\text{--}4}=+1.49$ at the MAP. To each catalogue we fit the same two models as in the first set, the thresholds in $\mathcal{F}_{\rm XUV}$ and in $P$ with the metallicity term and without age shifts. The difference of their log-likelihoods is at least $2.47$ in $36$ of the $500$ catalogues ($7\%$).

Taken together, the two sets show that a threshold in $P$ reproduces the observed difference only when an age dependence is added to it. This addition costs two parameters, whereas $\mathcal{F}_{\rm XUV}$ contains its age dependence with none. Fitted to the data, the threshold in $P$ with the age shifts exceeds the reference threshold by only $1.3$ in log-likelihood, so that the BIC favours the reference threshold by $10.1$. The two sets thus point to $\mathcal{F}_{\rm XUV}$ as the primary variable, with $P$ as its proxy across FGK hosts.

The third set tests whether the age shifts $\delta_m$ of Sect.~\ref{sec:age} can arise by chance. Its $200$ simulated catalogues are drawn from the reference threshold at its MAP, which has no age shifts. To each catalogue we fit the reference threshold without and with $\delta_m$. The gain in log-likelihood from $\delta_m$, which is $2.7$ for the data, is at least $2.7$ in $19$ of the $200$ catalogues (Sect.~\ref{sec:age} and Table~\ref{tab:evidence}). A gain as large as the observed one thus arises by chance in about $10\%$ of the catalogues, so that the data allow, but do not require, the age shifts. The fitted $\delta_{1\text{--}2}$ and $\delta_{2\text{--}4}$ reach their observed values in $2$ and $8$ of the $200$ catalogues, and both in $2$. The other comparisons of Table~\ref{tab:evidence} are tested in the same way, with $200$ catalogues simulated from the model listed second in each comparison, at its MAP.

The fourth set has one simulated catalogue for each of $300$ samples of the posterior of the reference threshold, and is not refitted. It gives the expected counted fractions and radius distributions of Figs.~\ref{fig:ppc} and~\ref{fig:radius1d}, and the frequencies with which discrepancies as large as the observed ones arise by chance, which are given in Appendix~\ref{app:tests}.

%% file: sections/D_planet_masses.tex
\section{The planet masses}\label{app:mass}

\subsection{Sample}
Among the planets of the extended sample $\mathcal{E}$ of Paper~II with radii of $0.5$--$4\,R_\oplus$, $155$ have a mass or a minimum mass in the catalogue of Paper~II and an implied density of $0.05$--$15$\,g\,cm$^{-3}$. The planets are classified as SEs or SNs according to the valley locus of Paper~II at their period, Eq.~\eqref{eq:locus}. We exclude the $34$ planets whose mass comes only from \citet{HaddenLithwick2014} or \citet{Xie2014}. These are nominal masses, inferred from transit-timing variations under the assumption of zero free eccentricity, which equal the true masses only if the eccentricities are small \citep{HaddenLithwick2014}. There remain $121$ planets, $29$ of them classified as SEs. Of their masses, $38$ have reliability tier $\le2$ and $83$ tier $3$, and $9$ are minimum masses from radial velocities, which equal the true masses to within a few per cent since these planets transit. The fluence $\mathcal{F}_Z$ (Eq.~\eqref{eq:fz}) uses $\mathcal{F}_{\rm XUV}$ at the adopted age of the host and $\xi_Z=-1.17$, the posterior median of the reference threshold. For the $18$ planets for which Paper~I gives only a limit on the host age, $\mathcal{F}_{\rm XUV}$ is evaluated at that limit.

\subsection{Model}
The model of Sect.~\ref{sec:massobs} describes how the classification of the $121$ planets, as SEs or SNs by the valley locus, depends on their fluence and mass. It has three parts.

First, the probability that planet $i$ is classified as a SE is a logistic function of its fluence $\mathcal{F}_{Z,i}$ (Eq.~\eqref{eq:fz}) and of its true mass,
\begin{equation}\label{eq:massmodel}
{\rm logit}\,{\Pr}_i({\rm SE}\mid z_i)=c_0+c_1\log_{10}\frac{\mathcal{F}_{Z,i}}{10^{21}\,{\rm erg\,cm^{-2}}}+c_2\,z_i,
\end{equation}
where $z_i$ is the true $\log_{10}(M_p/M_\oplus)$ of the planet. This probability differs from the SE weight $w_i$ of Eq.~\eqref{eq:law}, which weights the SE component of the mixture model of the radii (Sect.~\ref{sec:model}).

Second, the true masses of the planets at a given fluence follow a normal distribution,
\begin{equation}\label{eq:massdist}
z_i\sim\mathcal{N}\bigl(\bar z_i,\,\sigma_{M_p}^2\bigr),
\end{equation}
with mean
\begin{equation}\label{eq:zbar}
\bar z_i=b_0+b_1\log_{10}\frac{\mathcal{F}_{Z,i}}{10^{21}\,{\rm erg\,cm^{-2}}}.
\end{equation}
The dispersion $\sigma_{M_p}$ of Eq.~\eqref{eq:massdist} is the spread of the planet masses at fixed fluence of Sect.~\ref{sec:masswidth}.

Third, the likelihood of the measured mass $M_{p,i}^{\rm obs}$ is a split-normal function of the true mass, in linear mass, with the reported uncertainties. A large part of the masses are poorly known: $43$ of the $121$ are less than three times their lower uncertainty (Sect.~\ref{sec:massobs}). The likelihood of each planet is therefore integrated over its true mass,
\begin{equation}\label{eq:masslike}
\mathcal{L}_i=\int{\Pr}_i(k_i\mid z_i)\,\mathcal{N}\bigl(z_i;\bar z_i,\sigma_{M_p}^2\bigr)\,p\bigl(M_{p,i}^{\rm obs}\mid z_i\bigr)\,{\rm d}z_i.
\end{equation}
Here $k_i$ is the observed class of the planet, SE or SN, with
\begin{equation}\label{eq:prsn}
{\Pr}_i({\rm SN}\mid z_i)=1-{\Pr}_i({\rm SE}\mid z_i).
\end{equation}
In Eq.~\eqref{eq:masslike}, $p(M_{p,i}^{\rm obs}\mid z_i)$ is the split-normal likelihood of the measured mass, normalised as a density in $M_{p,i}^{\rm obs}$. The integral is computed on a grid of $450$ values of $z_i$ from $-1.5$ to $3.2$.

In Eqs.~\eqref{eq:massmodel} and~\eqref{eq:zbar}, $\mathcal{F}_{Z,i}$ is normalised to $10^{21}$\,erg\,cm$^{-2}$, close to the median of the sample, $10^{20.83}$\,erg\,cm$^{-2}$. The six parameters, $c_0$, $c_1$, $c_2$, $b_0$, $b_1$, and $\sigma_{M_p}$, are sampled with \texttt{emcee}. The priors are uniform in $c_0$, $c_1$, and $c_2$ between $-30$ and $30$, in $b_0$ between $-1$ and $3$, in $b_1$ between $-3$ and $3$, and in $\ln\sigma_{M_p}$ for $\sigma_{M_p}$ between $0.03$ and $3$\,dex. These ranges are much wider than the posteriors. Each chain has $24$ walkers and $10\,000$ steps, and the first third is discarded. The retained part is more than $55$ times the longest autocorrelation time.

The boundary of Eq.~\eqref{eq:massboundary} is where the right-hand side of Eq.~\eqref{eq:massmodel} vanishes. Its slope and its value at $2.5\,M_\oplus$ are, respectively:
\begin{equation}\label{eq:aMxM}
a_M=-\frac{c_2}{c_1},\qquad h_M=21-\frac{c_0}{c_1}+a_M\log_{10}2.5.
\end{equation}

\subsection{Width of the threshold}
The width $\sigma_{\mathcal{F}}$ of the reference threshold has two contributions. The first comes from the planets themselves. Each planet belongs to the SE component above a critical value of $\log_{10}\mathcal{F}_Z$, which differs from planet to planet (Sect.~\ref{sec:thrforms}). We call $\sigma_{\rm thr}$ the dispersion of these critical values. The second contribution is the scatter $\sigma_{\rm rot}=0.17$\,dex of $\mathcal{F}_{\rm XUV}$ among hosts of the same mass, due to their unknown initial rotation (Appendix~\ref{app:axes}). 
If both are normal, they add in quadrature:
\begin{equation}\label{eq:sigmaF}
\sigma_{\mathcal{F}}^2=\sigma_{\rm thr}^2+\sigma_{\rm rot}^2.
\end{equation}
With this relation, the posterior of $\sigma_{\mathcal{F}}$ gives $\sigma_{\rm thr}=0.27\pm0.14$\,dex, with an 84th percentile of $0.40$\,dex (Sect.~\ref{sec:masswidth}).

If the critical value of each planet grows with its core mass as $M_{\rm c}^{a}$, a spread $\sigma_{M_{\rm c}}$ of $\log_{10}M_{\rm c}$ contributes $a\,\sigma_{M_{\rm c}}$ to $\sigma_{\rm thr}$. Other properties of the planets, such as the envelope mass fraction and the escape efficiency, also change their critical values at fixed core mass. Their contribution has a dispersion $\sigma_{\rm ot}$ and adds in quadrature,
\begin{equation}\label{eq:sigmathr}
\sigma_{\rm thr}^2=a^2\sigma_{M_{\rm c}}^2+\sigma_{\rm ot}^2\ge a^2\sigma_{M_{\rm c}}^2,
\end{equation}
so that
\begin{equation}\label{eq:mclimit}
\sigma_{M_{\rm c}}\le\frac{\sigma_{\rm thr}}{a}.
\end{equation}
With the 84th percentile of $\sigma_{\rm thr}$, this gives the limits of Sect.~\ref{sec:masswidth}, $0.34$ and $0.20$\,dex for $a=1.2$ and $2$, respectively. For the slope of the model of \citet{OwenWu2017}, $a=1.42$, the limit is $0.28$\,dex, equal to the dispersion in $\log_{10}M_{\rm c}$ of the core masses in that model, which follow a Rayleigh distribution of mode $3\,M_\oplus$. These limits depend little on the value of $\sigma_{\rm rot}$, which comes from the spread of the activity tracks of \citet{Johnstone2021}. Without its subtraction in Eq.~\eqref{eq:sigmaF}, they rise only to $0.37$ and $0.22$\,dex.

If the spread $\sigma_{M_{\rm c}}$ of the core masses equals the spread $\sigma_{M_p}$ of the planet masses (Eq.~\eqref{eq:massdist}), Eq.~\eqref{eq:mclimit} requires $\sigma_{M_p}\le\sigma_{\rm thr}/a$. For the $121$ planets, $a_M=2.3^{+0.8}_{-0.5}$ and $\sigma_{M_p}=0.35\pm0.03$\,dex (Sect.~\ref{sec:masswidth}). The sample is small and many masses have large relative uncertainties, so that these values are not stable when the fit is restricted to a subset of the masses. For the $97$ masses larger than twice their lower uncertainty, $21$ of them SEs, $a_M=1.8^{+0.7}_{-0.5}$ and $\sigma_{M_p}=0.29^{+0.03}_{-0.02}$\,dex. For the $78$ masses larger than three times their lower uncertainty, $14$ of them SEs, $a_M=1.4^{+0.6}_{-0.5}$ and $\sigma_{M_p}=0.28\pm0.03$\,dex. For the $38$ masses of tier $\le2$, $a_M=2.0^{+1.2}_{-0.8}$ and $\sigma_{M_p}=0.23^{+0.04}_{-0.03}$\,dex. For the $75$ planets whose $\log_{10}\mathcal{F}_Z$ lies within $0.5$\,dex of $y_{\rm c}$, $a_M=1.0^{+0.6}_{-0.4}$ and $\sigma_{M_p}=0.33^{+0.04}_{-0.03}$\,dex.

\subsection{Expected slopes}\label{app:massslopes}
If the critical values of the reference threshold (Sect.~\ref{sec:thrforms}) are set by energy-limited escape, a planet loses its envelope, of mass $M_{\rm env}$, when a constant fraction $\eta_{\rm esc}$ of the XUV energy it has absorbed equals the binding energy of the envelope. With $M_p\approx M_{\rm c}$, the XUV fluence at which this occurs, $\mathcal{F}_{\rm XUV,thr}$, satisfies
\begin{equation}\label{eq:budget}
\eta_{\rm esc}\,\mathcal{F}_{\rm XUV,thr}\,\pi R_{\rm abs}^2=\frac{GM_{\rm c}M_{\rm env}}{R_{\rm bind}},
\end{equation}
where $R_{\rm abs}$ is the radius at which the XUV radiation is absorbed and $R_{\rm bind}$ that at which the binding energy is evaluated. If the envelope fraction and the two radii scale with the core mass as
\begin{equation}\label{eq:budgetscaling}
\frac{M_{\rm env}}{M_{\rm c}}\propto M_{\rm c}^{n_{\rm env}},\qquad R_{\rm bind}\propto M_{\rm c}^{n_{\rm bind}},\qquad R_{\rm abs}\propto M_{\rm c}^{n_{\rm abs}},
\end{equation}
Eq.~\eqref{eq:budget} gives
\begin{equation}\label{eq:budgetslope}
\mathcal{F}_{\rm XUV,thr}\propto M_{\rm c}^{a},\qquad a=2+n_{\rm env}-n_{\rm bind}-2n_{\rm abs}.
\end{equation}
The slope thus depends on how the envelope fraction and the two radii scale with the core mass, and different assumptions give the range of $1.2$ to $2$ quoted in Sect.~\ref{sec:massobs}. For a rocky core, we adopt the mass--radius relation
\begin{equation}\label{eq:rcore}
R_{\rm c}=R_\oplus\Bigl(\frac{M_{\rm c}}{M_\oplus}\Bigr)^{0.27}
\end{equation}
\citep[Sect.~\ref{sec:massobs};][]{Valencia2006,Zeng2016}. With a fixed envelope fraction ($n_{\rm env}=0$) and both $R_{\rm bind}$ and $R_{\rm abs}$ proportional to $R_{\rm c}$ ($n_{\rm bind}=n_{\rm abs}=0.27$), the slope is $a=1.19$. With $R_{\rm abs}$ independent of the core mass ($n_{\rm abs}=0$), it is $a=1.73$. In the model of \citet{OwenWu2017}, the mass-loss time-scale is longest for $R_p\simeq2R_{\rm c}$. For this radius, and for cores with $R_{\rm c}\propto M_{\rm c}^{1/4}$, $a=1.42$. In the analytical model of \citet{Mordasini2020}, the most massive planet that loses its whole envelope has $M_{\rm c}\propto\mathcal{F}_{\rm XUV,thr}^{1/2}$, so that $a=2$, independently of how the envelope mass scales with the core mass.

To estimate the order of magnitude of $\mathcal{F}_{\rm XUV,thr}$ (Sect.~\ref{sec:discmass}), we take a rocky core of mass $M_{\rm c}=5\,M_\oplus$ and radius from Eq.~\eqref{eq:rcore}, with $M_{\rm env}=0.02\,M_{\rm c}$, $R_{\rm bind}=R_{\rm c}$, $R_{\rm abs}=2.5\,R_\oplus$, and $\eta_{\rm esc}=0.1$. Equation~\eqref{eq:budget} then gives $\mathcal{F}_{\rm XUV,thr}=10^{21.2}$\,erg\,cm$^{-2}$.

\subsection{Cosmic shoreline}\label{app:shoreline}
Figure~\ref{fig:planetmass} compares the boundary of Eq.~\eqref{eq:massboundary} with two estimates of the cosmic shoreline (Sect.~\ref{sec:massobs}), $\mathcal{F}_{\rm XUV}/\mathcal{F}_\oplus\propto v_{\rm esc}^{4}$ \citep{ZahnleCatling2017} and $\mathcal{F}_{\rm XUV}/\mathcal{F}_\oplus\propto v_{\rm esc}^{5.77}$ \citep{MeniGallardo2026}, where $\mathcal{F}_\oplus$ is the XUV fluence of the Earth. \citet{ZahnleCatling2017} draw their relation without a fit and give no zero point; we fix it at Mars, which lies on their line. \citet{MeniGallardo2026} fix both the slope and the zero point with Mars and four exoplanets with tentative atmospheric detections. Neither relation has a quoted uncertainty.

We compute $\mathcal{F}_\oplus$ for the Sun (with $1\,M_\odot$ and $4.57$\,Gyr), with the same $R_{\rm X}$ as for the hosts (Appendix~\ref{app:axes}) and the orbital period of the Earth, so that $\mathcal{F}_\oplus$ and the $\mathcal{F}_{\rm XUV}$ of the planets come from the same activity model. This gives $\log_{10}(\mathcal{F}_\oplus/{\rm erg\,cm^{-2}})=18.94$. Both literature estimates of the cosmic shoreline give $\mathcal{F}_{\rm XUV}$, which equals the $\mathcal{F}_Z$ of Fig.~\ref{fig:planetmass} for hosts with $\Delta_i=0$. To draw them in the plane of Fig.~\ref{fig:planetmass}, we write the escape velocity as a function of $M_p$ alone, with the rocky radius of Eq.~\eqref{eq:rcore}, that of a planet without envelope:
\begin{equation}\label{eq:vesc}
v_{\rm esc}=\Bigl(\frac{2GM_p}{R_{\rm c}}\Bigr)^{1/2}=11.19\,{\rm km\,s^{-1}}\Bigl(\frac{M_p}{M_\oplus}\Bigr)^{0.365},
\end{equation}
where $11.19$\,km\,s$^{-1}$ is the escape velocity of the Earth.

With Eq.~\eqref{eq:vesc}, the estimate of \citet{ZahnleCatling2017} gives $\mathcal{F}_{\rm XUV}\propto M_p^{1.46}$ and that of \citet{MeniGallardo2026} gives $\mathcal{F}_{\rm XUV}\propto M_p^{2.11}$, while the boundary of Eq.~\eqref{eq:massboundary} has slope $a_M=2.3^{+0.8}_{-0.5}$. At $2.5\,M_\oplus$ the two estimates lie at $\log_{10}(\mathcal{F}_{\rm XUV}/{\rm erg\,cm^{-2}})=20.5$ and $21.5$, on either side of the $h_M=21.00^{+0.16}_{-0.14}$ of Eq.~\eqref{eq:massboundary}.

These positions depend on the radius used in Eq.~\eqref{eq:vesc}. At $2.5\,M_\oplus$, replacing the radius of a rocky planet given by Eq.~\eqref{eq:rcore}, $1.28\,R_\oplus$, with the median radius of the SN component, $2.46\,R_\oplus$ (Sect.~\ref{sec:ppc}), lowers the $\log_{10}\mathcal{F}_{\rm XUV}$ of the two estimates to $20.0$ and $20.7$, both below $h_M$.

%% file: sections/E_closed_form.tex
\section{The reference threshold in closed form}\label{app:closedform}

This appendix gives an approximation of the reference threshold that needs neither the rotational model nor the turnover times. It gives the SE weight of a planet from its observables and those of its host, with the parameters of Table~\ref{tab:params}. It holds for hosts and orbits like those of $\mathcal{D}_3$: FGK hosts of $0.54$ to $1.23\,M_\odot$ with ages of $0.3$ to $4$\,Gyr, orbital periods up to $100$\,d, and [Fe/H] between $-0.22$ and $+0.24$ (5th--95th percentiles). The background fraction $w_{\rm b}$ refers to planets of $0.5$ to $4\,R_\oplus$ with the probabilities of detection and confirmation of the \textit{Kepler} planets of $\mathcal{D}_3$ (Sect.~\ref{sec:model}). The inputs are, for the planet, its orbital period $P$; for the host, its mass $M_\star$, radius $R_\star$, effective temperature $T_{\rm eff}$, metallicity [Fe/H], and gyrochronological age $t$; and, for the system, the confirmed-sibling indicator $g_i$, which marks the planets with at least one other confirmed planet around the same host (Sect.~\ref{sec:sample}). The bolometric luminosity $L$ and the semi-major axis $a_p$ are
\begin{equation}\label{eq:lbol_closed}
L=\bigl(R_\star/R_\odot\bigr)^{2}\,\bigl(T_{\rm eff}/5772\,\mathrm{K}\bigr)^{4}\,L_\odot,
\end{equation}
\begin{equation}\label{eq:sma_closed}
\bigl(a_p/{\rm au}\bigr)^{3}=\bigl(M_\star/M_\odot\bigr)\,\bigl(P/365.25\,{\rm d}\bigr)^{2}.
\end{equation}
We approximate the integrand of Eq.~\eqref{eq:fx}, with the $R_{\rm X}$ of Appendix~\ref{app:axes}, as
\begin{equation}\label{eq:integrand_closed}
R_{\rm X}(t')\,\frac{L_{\rm M}(t')}{L_{\rm M}(t)}\simeq\tilde R_{\rm X}(t')\,\Bigl(\frac{t'}{t}\Bigr)^{\tilde\beta},
\end{equation}
where $\tilde R_{\rm X}$ has the two-segment form of Eq.~\eqref{eq:rx},
\begin{equation}\label{eq:rxtilde_closed}
\tilde R_{\rm X}(t')=\begin{cases}\tilde R_{\rm sat}, & t'\le\tilde\tau_{\rm sat},\\ \tilde R_{\rm sat}\,(t'/\tilde\tau_{\rm sat})^{-\tilde\alpha}, & t'>\tilde\tau_{\rm sat}.\end{cases}
\end{equation}
The four parameters $\tilde R_{\rm sat}$, $\tilde\tau_{\rm sat}$, $\tilde\alpha$, and $\tilde\beta$ are parabolas in the host mass, through
\begin{equation}\label{eq:zeta_closed}
\zeta=M_\star/M_\odot-1.
\end{equation}
We fit the three coefficients of each parabola, twelve in all, to this $R_{\rm X}$, as described below, and obtain:
\begin{equation}\label{eq:rsat_closed}
\log_{10}\tilde R_{\rm sat}=-3.191-0.186\,\zeta+2.112\,\zeta^{2},
\end{equation}
\begin{equation}\label{eq:tausat_closed}
\log_{10}(\tilde\tau_{\rm sat}/{\rm Gyr})=-2.007-1.671\,\zeta-0.334\,\zeta^{2},
\end{equation}
\begin{equation}\label{eq:alpha_closed}
\tilde\alpha=0.820+0.891\,\zeta+0.308\,\zeta^{2},
\end{equation}
\begin{equation}\label{eq:beta_closed}
\tilde\beta=-0.077-1.111\,\zeta+7.487\,\zeta^{2}.
\end{equation}
With these power laws, the integrals in the accumulated XUV fluence can be solved analytically. The two extreme-ultraviolet parts of Eq.~\eqref{eq:euv} scale with the powers $q_1=0.681$ and $q_2=0.681\times0.920$ of the X-ray surface flux $f_{\rm X}$ (Eq.~\eqref{eq:fxsurf}). The XUV fluence thus contains time integrals of three powers of the approximated integrand of Eq.~\eqref{eq:integrand_closed}, $p=1$ for the X-ray part and $p=q_1$ and $p=q_2$ for the two extreme-ultraviolet parts. For a power $p$, the integral is $\tilde J_p$:
\begin{equation}\label{eq:jp_closed}
\begin{aligned}
\tilde J_p(t)&=\int_0^{t}\Bigl[\tilde R_{\rm X}(t')\Bigl(\frac{t'}{t}\Bigr)^{\tilde\beta}\Bigr]^{p}\,{\rm d}t'\\
&=\frac{\tilde R_{\rm sat}^{p}}{t^{p\tilde\beta}}\Biggl[\frac{\min(t,\tilde\tau_{\rm sat})^{1+p\tilde\beta}}{1+p\tilde\beta}\\
&\qquad\quad+\tilde\tau_{\rm sat}^{p\tilde\alpha}\,\frac{\max(t,\tilde\tau_{\rm sat})^{1+p\tilde\beta-p\tilde\alpha}-\tilde\tau_{\rm sat}^{1+p\tilde\beta-p\tilde\alpha}}{1+p\tilde\beta-p\tilde\alpha}\Biggr].
\end{aligned}
\end{equation}
The second term in square brackets is the contribution of the declining phase after the end of the saturated phase. When $1+p\tilde\beta-p\tilde\alpha=0$, which occurs near $1.07$ and $1.21\,M_\odot$ for $p=1$, it is replaced by its limit,
\begin{equation}\label{eq:jp_limit}
\tilde\tau_{\rm sat}^{p\tilde\alpha}\,\ln\frac{\max(t,\tilde\tau_{\rm sat})}{\tilde\tau_{\rm sat}}.
\end{equation}
The X-ray part and the two extreme-ultraviolet parts then give the approximate XUV fluence
\begin{equation}\label{eq:fluence_closed}
\begin{aligned}
\tilde{\mathcal{F}}_{\rm XUV}(t)=\frac{1}{4\pi a_p^{2}}\Bigl[&L\,\tilde J_1(t)+110\,A^{1-q_1}L^{q_1}\tilde J_{q_1}(t)\\
&+0.924\times110^{0.920}\,A^{1-q_2}L^{q_2}\tilde J_{q_2}(t)\Bigr].
\end{aligned}
\end{equation}
Here $L$ is in erg\,s$^{-1}$ ($L_\odot=3.828\times10^{33}$\,erg\,s$^{-1}$), $A=4\pi R_\star^2$ is the stellar surface area in cm$^2$ ($R_\odot=6.957\times10^{10}$\,cm), $a_p$ is in cm (1\,au $=1.496\times10^{13}$\,cm), $t$ and $\tilde\tau_{\rm sat}$ are in Gyr, and $\tilde J_p$ is converted from Gyr to s (1\,Gyr $=3.156\times10^{16}$\,s).

The twelve coefficients of Eqs.~\eqref{eq:rsat_closed}--\eqref{eq:beta_closed} are fitted jointly, by least squares in $\log_{10}\tilde J_p$, to the integrals of Eq.~\eqref{eq:jp_closed} evaluated with the $R_{\rm X}(t')\,L_{\rm M}(t')/L_{\rm M}(t)$ of the tracks in place of its approximation, as in the numerical $\mathcal{F}_{\rm XUV}$. The four parameters are therefore effective. They reproduce the integrals, not $R_{\rm X}(t)$ or $L_{\rm M}(t)$ themselves, and should not be compared with those measured from the observed X-ray emission. The fit uses the three exponents of Eq.~\eqref{eq:fluence_closed}, the masses from $0.50$ to $1.25\,M_\odot$ (Appendix~\ref{app:axes}), and $80$ ages spaced evenly in logarithm from $1$\,Myr to $10$\,Gyr or to the end of the track, whichever is earlier.

The closed form reproduces these integrals to within $0.20$\,dex between $0.1$ and $6$\,Gyr. At older ages the deviation grows for hosts above $0.85\,M_\odot$, and reaches $0.39$\,dex at $10$\,Gyr for $1\,M_\odot$. Over $\mathcal{D}_3$, $\tilde{\mathcal{F}}_{\rm XUV}$ differs from the $\mathcal{F}_{\rm XUV}$ of Sect.~\ref{sec:axes}, at the age of each host, by a median of $-0.04$\,dex, with a dispersion of $0.03$\,dex and at most $0.13$\,dex.

With $\Delta_i={\rm [Fe/H]}_i-{\rm [Fe/H]}_0$ as in Eq.~\eqref{eq:law} and $\Phi$ the standard normal cumulative distribution function, the SE weight is approximated, directly in terms of these inputs, by $\tilde w_i$:
\begin{equation}\label{eq:law_closed}
{\rm logit}\,\tilde w_i={\rm logit}\bigl[w_{\rm b}+(1-w_{\rm b})\,\Phi(\tilde\Lambda_i)\bigr]+\gamma\,g_i,
\end{equation}
\begin{equation}\label{eq:lambda_closed}
\tilde\Lambda_i=\sigma_{\mathcal{F}}^{-1}\Bigl[\log_{10}\tilde{\mathcal{F}}_{\rm XUV}\bigl(P,M_\star,R_\star,T_{\rm eff},t\bigr)-y_{\rm c}+\xi_Z\,\Delta_i\Bigr].
\end{equation}
With the posterior medians of Table~\ref{tab:params}, $\tilde w_i$ differs from the weight $w_i$ of Eq.~\eqref{eq:law} at the age of each host of $\mathcal{D}_3$ by a median of $-0.004$ and by at most $0.08$. Refitted at the MAP, with the age term $\mathcal{H}$ of the confirmation probability fixed at zero (Appendix~\ref{app:fits}) and with $\tilde{\mathcal{F}}_{\rm XUV}$ in place of $\mathcal{F}_{\rm XUV}$, the reference threshold reaches a log-likelihood of $423.02$, compared with $423.38$ with $\mathcal{F}_{\rm XUV}$.